\documentclass{article}

\usepackage{arxiv}
\usepackage[utf8]{inputenc} 
\usepackage[T1]{fontenc}    
\usepackage{hyperref}       
\usepackage{url}            
\usepackage{microtype}      
\usepackage{lipsum}
\usepackage{graphicx}       
\usepackage[british]{babel}
\usepackage{csquotes}

\usepackage[backend=biber,citestyle=numeric,style=numeric-comp,sorting=none,sortcites=true,maxbibnames=6,minbibnames=6,alldates=long,autopunct=true,language=british]{biblatex}

\title{The Risk to Population Health Equity Posed by Automated Decision Systems}

\author{
  Mitchell~Burger\\
  School of Population Health\\
  UNSW Sydney\\
  Sydney NSW Australia\\
  \texttt{m.burger@student.unsw.edu.au}\\
 }

\begin{document}
\maketitle

\begin{abstract}

\textbf{Introduction: }Artificial intelligence is already ubiquitous, and is increasingly being used to autonomously make ever more consequential decisions. Amid scrutiny of the use of narrow artificial intelligence (which means performing singular tasks) and automated decision systems in sensitive domains such as policing and education, there is concern about impacts on equity and justice. However, there has been relatively little research into the existing and possible consequences for population health and equity of their use in medicine and public health.

\textbf{Method: }A narrative review of scholarly and grey literature was undertaken using a hermeneutic approach. The review explores current and future uses of narrow artificial intelligence and automated decision systems in medicine and public health, issues that have emerged, and longer-term implications for population health equity.

\textbf{Results: }Accounts reveal a tremendous expectation on artificial intelligence to transform medical and public health practices, especially regarding precision medicine and precision public health. Increasingly frequent and prominent demonstrations of narrow artificial intelligence capability – particularly in diagnostic decision making, risk prediction, and disease surveillance – are stimulating a rapid expansion in adoption, spurred by the COVID-19 pandemic. Automated decisions being made about disease detection, diagnosis, treatment, and health funding allocation have significant consequences for individual and population health and wellbeing. Meanwhile, it is evident that hazards including bias, incontestability, and erosion of privacy have emerged in sensitive domains where narrow artificial intelligence and automated decision systems are in common use, such as criminal justice and education. Reports of issues arising from their use in health are already appearing.

\textbf{Discussion: }As the use of automated decision systems expands in medicine and public health, it is probable that these hazards will manifest more widely. Bias, incontestability, and privacy erosion give rise to mechanisms by which existing social, economic and health disparities are perpetuated and amplified. Consequently, there is a significant risk that use of automated decision systems in health will exacerbate existing population health inequities.

\textbf{Conclusion: }The industrial scale and rapidity with which automated decision systems can be applied to whole populations heightens the risk to population health equity. It is incumbent on health practitioners and policy makers to explore the potential implications of using automated decision systems, to ensure the use of artificial intelligence promotes population health and equity.

\end{abstract}

\keywords{artificial intelligence \and AI \and automated decision systems \and public health \and precision public health \and population health \and equity}

\section{Introduction}
There is tremendous hype surrounding the future of artificial intelligence\footnote{The UK Parliament’s House of Lords Select Committee on Artificial Intelligence \parencite*{HouseLords18Ready} wryly observed that “the debate around exactly what is, and is not, artificial intelligence, would merit a study of its own”. For the purposes of this review the practical definition of artificial intelligence specified by the House of Lords Select Committee is used: ‘Technologies with the ability to perform tasks that would otherwise require human intelligence, such as visual perception, speech recognition, and language translation.’ (p. 20)} (AI) in health – particularly in medicine \parencite{Coiera18,Maddox19,IqbalU20,Wilkinson20}, and increasingly in public health \parencite{Chowkwanyun18,TaylorRobinson18,WongZ19,Khoury20,Brockmann20}. The singular, global impact of AI on population health has been called out by the World Health Organization, who proclaim that “more human lives will be touched by health information technology than any other technology, ever.” \parencite{WHO18BigData} While predictions vary about the extent to which AI will actually revolutionise medicine and public health practices in the shorter term \parencite[as discussed by, for example,][]{Coiera18,Darcy16,Mukherjee17,TaylorRobinson18,Wilkinson20,Yu18}, scholars have identified the longer-term potential of AI to reduce global health inequalities \parencite{Alami20,Mehta20,Schwalbe20}. Even the United Nations Secretary-General has highlighted the potential of AI to advance human welfare, but has also emphasised its potential to widen inequality and increase violence \parencite{UN18SecGeneralStrat}.

The hype about the impact of artificial intelligence in medicine and public health comes at a time of unprecedented global interest in AI, wherein the potential future impacts of AI are being extensively analysed and discussed, including on the pages of this journal \parencite[e.g.,][]{Dolley18,Weeramanthri18,Triberti20,Lefevre20}. It has been common for the future risks of AI to predominate, especially those relating to automation \parencite[e.g.,][]{Berg18,DawsonN18,Frontier18,Furman16,Morcom18,Schwab16,RoyalSociety18}, autonomous weapons \parencite[e.g.,][]{DawsonN18,Future15,Schneier19Data}, and superintelligence \parencite[e.g.,][]{Bostrom14,Bridle18,Brundage15,Pamlin15,Tegmark17}.

However, while the idea of artificial intelligence may still readily conjure science fiction dreams and nightmares in popular imagination, the reality is that AI is here already \parencite{Campolo17,Rahwan19}. Meredith Whittaker and colleagues at the AI Now Institute have stated: “The rapid deployment of AI and related systems in everyday life is not a concern for the future—it is already here, with no signs of slowing down.” \parencite{Whittaker18Report} Narrow artificial intelligence\footnote{‘Narrow’ AI (also referred to as ‘weak’ AI) generally refers to an agent that undertakes a specific, singular task; typically in a way that does not generalise to other tasks or domains \parencite{Russell10}. For context, higher grades of AI are: ‘artificial general intelligence’, which is also referred to as ‘strong AI’, and is taken to mean an agent performing tasks at least as well as humans across many or all domains \parencite{Bughin17,DawsonN18}. ‘Superintelligence’ is performance that exceeds human intelligence across all domains \parencite{Bostrom14}.} and predictive algorithms suffuse society – they are woven into the fabric of our daily lives \parencite{Amoore09,Mackenzie15,Zook17}; mediating “our social, cultural, economic and political interactions” \parencite{Rahwan19}. In this way, AI is already ubiquitous \parencite{Campolo17,Lavigne19,Schneier19Supply}, often in very mundane forms in everyday technologies \parencite{Mackenzie15} – smart phones, online advertising, social media, home assistants, recommendation engines for music and video, online dating, autopilots, and customer support chatbots.

‘Automated decision systems’ are a particular mode of implementing AI which use classifications and predictions produced by expert systems or machine learning algorithms to make decisions \textit{autonomously}.\footnote{It is difficult to more precisely define ‘automated decision systems’ because they are specific to context and purpose \parencite{Reisman18AlgoImpactAssess}. For example, Reisman and colleagues define predictive policing automated decision systems as “any systems, tools, or algorithms that attempt to predict crime trends and recommend the allocation of policing resources” (p. 13). A definition adapted for health could be: \textit{any systems, tools, or algorithms that attempt to predict individual or population health trends or states, and decide the allocation of health resources or specific interventions}. This definition distinguishes automated decision systems from computerised decision \textit{support} systems \parencite[e.g.,][]{Catho21}, which are intended to guide clinician-in-the-loop decision-making. Automated decision systems would also be a higher risk categorisation in the FDA's emerging regulatory framework for artificial intelligence or machine learning-based software-as-a-medical device \parencite{FDA20}.} Such systems are already prevalent in a broad range of sectors, including loan and credit card applications, algorithmic trading, drone warfare, immigration, criminal justice, policing, job applications, education, university entry, utilities network management, and social welfare \parencite{Angwin16,Campolo17,Knight17,Lecher18,Rahwan19,Reisman18AlgoImpactAssess,Laplante20}. Their use by governments and corporations is rapidly expanding into ever more consequential and sensitive domains \parencite{Rahwan19,Whittaker18Report}. And what is particularly insidious about the expanding use of automated decision making is the \textit{invisibility} of their proliferation \parencite{Campolo17}. Crawford and colleagues write: “In many cases, people are unaware that a machine, and not a human process, is making life-defining decisions.” \parencite*[p. 23]{Crawford16Report} In addition, people tend to become rapidly habituated to advances in AI performance, leading to creeping normalisation. Contributing to this normalisation is the tendency for AI to have an ever-evolving definition as \textit{what is not yet possible} \parencite{Kurzweil05}.\footnote{Indeed, Tesler’s theorem – attributed to the computer scientist Larry Tesler – is that: “AI is whatever hasn’t been done yet.” \parencite[p. 7]{Bughin17}. Pioneering AI researcher John McCarthy also had a similar saying: “As soon as it works, no one calls it AI anymore.” \parencite{Meyer11}.} Mundanity, invisibility, and habituation are enabling automated decision systems to proliferate unseen.

As artificial intelligence takes on more and more responsibility for consequential decisions, fundamental questions of rights, fairness and equity arise \parencite{Crawford16Report,AHRC2019report,Baeroe20,Fosso21}. There is substantiated concern about the impact AI is already having on equity and justice, with mounting evidence that AI systems can perpetuate, entrench and amplify existing discrimination and inequality \parencite[e.g.,][]{AINow18YearReview,Campolo17,Crawford16Report,Eubanks17,UN18ReportSpecialRap}. This is prompting widespread debate about where and when AI and automated decision systems can be used \parencite[e.g.,][]{Campolo17,Walsh17ABC,Goodman20}. However, while high-level warnings about medium and long-term risks and societal impacts are widespread, the effects of the unseen proliferation of narrow AI and automated decision systems in sensitive domains have only relatively recently started to be closely scrutinised. In health – compared to domains such as criminal justice, policing and autonomous warfare – there has been less in-depth analysis of the consequences for population health and population health equity of the use of narrow AI and automated decision systems. Acknowledging the conceptual 'fuzziness' \parencite{Krieger12}, for our purposes ‘population health’ can be taken to mean the ‘collective health’ of populations, drawing on Rose's conception that “healthiness is a characteristic of the population as a whole and not simply of its individual members.” \parencite[p. 95]{Rose08} Population health \textit{equity} on the other hand is a political concept requiring judgement based on concepts of social justice as to whether measurable differences in health (inequalities) are unjust or avoidable \parencite{Kawachi02}. Key in this context is the differential impact of social determinants of health (Marmot, 2005), and how these determinants are accounted for (or not) in data and automated decision systems \parencite{Galea20}.

While accounts in medical and public health literature have posed the question about the effect on equity, and identified risks \parencite[e.g.,][]{Bauer21,Fenech18,Smith20}, analysis to date has focused primarily on the ethical and legal implications of medical applications for individuals \parencite[e.g.,][]{Carter20}. The potential impact on population health equity of issues known to have emerged widely in other sensitive domains, and the mechanisms by which they are likely act, therefore require further investigation.

\section{Method}
This review aims to begin to address this evidence gap regarding the use of automated decision systems in public health by reviewing the current state of adoption of AI in health, drawing together detailed evidence of issues arising in other sensitive domains, and specifically considering the implications for population health equity, including the mechanisms by which population health equity may be affected. The specific research questions are:
\begin{enumerate}
\item Broadly, what is the current state of adoption of narrow AI and automated decision systems in medicine and public health? And how are they expected to be used in the future?
\item What key issues of relevance to equity have emerged in the application of narrow AI and automated decision making in other sensitive domains? Is there evidence of these issues emerging in medicine and/or public health applications?
\item What are the possible implications and risks for population health equity?
\end{enumerate}

To address these questions, a narrative review \parencite{Ferrari2015} using a hermeneutic approach has been undertaken. A hermeneutic review involves an iterative process of developing understanding through cycles of search and acquisition of literature, together with iterative analysis and interpretation \parencite{Boell2014}. It is an approach that it is suitable for questions requiring clarification and insight which cover diverse and dynamic bodies of scholarly and grey literature \parencite{Boell2014,Greenhalgh18}. Furthermore, this approach is consistent with Galea and colleagues’ \parencite{Galea20} call for transdisciplinary synthesis, in that the review draws on a range of literatures, including technical, public health, policy, and sociology.

Initially, literature was gathered by searching Scopus, PubMed, Web of Science, and IEEE Xplore databases, and arXiv, medRxiv, and bioRxiv pre-print servers using the search terms ‘artificial intelligence’, ‘machine learning’, or ‘big data’, in combination with ‘public health’ or ‘epidemiology’. Searches were also conducted using the search terms ‘artificial intelligence’ or ‘machine learning’, together with ‘bias’ or ‘privacy’. Grey literature was sourced using Google Scholar, Hacker News\footnote{\url{https://news.ycombinator.com/}}, WHO IRIS, United Nations Official Document System, and World Economic Forum Reports. Relevant articles were selected by scanning titles and abstracts, yielding 240 references. Citation tracking was then used to identify additional sources as analysis proceeded, in keeping with the hermeneutic approach. Mapping, classification, and thematic analysis was undertaken iteratively using NVivo 12 Pro qualitative analysis software.

\section{Results}
\subsection{Current state of adoption in health, and expected future uses}
While the use of narrow artificial intelligence and automated decision systems is already widespread in sectors such as finance, policing, and criminal justice; the health sector has a comparatively low – but rapidly expanding – level of adoption \parencite{Bughin17,Lancet19,Zandi19}. In 2017, an American independent scientific advisory group JASON\footnote{See Federation of American Scientists \parencite*{FederationJASON19}} described the state of adoption of AI in the health sector generally as being at an exploratory phase: “AI is beginning to play a growing role in transformative changes now underway in both health and health care, in and out of the clinical setting. At present the extent of the opportunities and limitations is just being explored.” \parencite[p. 1]{Derrington17} In the ensuing years the state of adoption of AI in health has advanced markedly, with increasing and sometimes hyperbolic reports that AI is now starting to replace doctors \parencite[e.g.,][]{LePage19,Beckman20,Shuaib20}. The most active areas of application are use of machine learning for diagnostic support, for example in medical imaging interpretation and multivariate risk prediction.\footnote{Machine learning is a sub-branch of artificial intelligence research and practice – for definitions of machine learning and deep learning see for example Brooks \parencite*{Brooks17Machine}, Russell and Norvig  \parencite*{Russell10}, and LeCun et al. \parencite*{LeCun15}. Challen and colleagues \parencite*{Challen19} plotted the current state of adoption of machine learning in healthcare as at 2019, and projected future applications in terms of increasing levels of automation. Similarly, Ching and colleagues \parencite*{Ching18} provide an excellent and comprehensive survey of the state of adoption of deep learning in medicine.}

In medicine, there have been prominent and increasingly frequent demonstrations of the capability of AI – in particular machine learning – to perform diagnostics using medical images with the same performance levels as experienced clinicians \parencite{Coiera18,Derrington17,Yasaka18,Liu19,Koopman20,AdamsS21}. High-profile examples include: automated detection of key trauma and stroke indications in head CT scans \parencite[in \textit{The Lancet}]{Chilamkurthy18}; automated classification of skin cancer at dermatologist-level \parencite[in \textit{Nature}]{Esteva17}; detection of diabetic retinopathy \parencite[in \textit{JAMA}]{Gulshan16}; and automated classification of abnormalities in chest radiographs \parencite[in \textit{PLOS Medicine}]{Rajpurkar18}. A systematic review published in \textit{The Lancet Digital Health} found the “diagnostic performance of deep learning models to be equivalent to that of health-care professionals”, although concerns were raised about the prevalence of poor reporting in deep learning studies \parencite{Liu19}. Since 2018 there has been an acceleration in US Food and Drug Administration (FDA) approvals of AI algorithms \parencite{Mesko19}. As a further indicator of the accelerating advance of AI adoption, AI-based tools have been widely implemented in response to the COVID-19 pandemic \parencite[e.g.,][]{Dorr20,Haase21,Horgan20,IqbalS21,Ozturk20,Peek20,Saniotis20,Scott20}. However, to date there has been little high-quality evidence establishing the efficacy of these tools in practice \parencite{Cresswell20}. Future applications include “using artificial intelligence and machine learning to support the integration of genomic information into health care systems” \parencite[p. 22]{Williamson18}, so as to enable personalised drug protocols, precision prevention \parencite{Meagher17}, and early diagnosis of rare childhood diseases \parencite{Wright18}.

In public health accounts, AI is typically regarded with cautious optimism as having the potential to re-envisage and transform public health practices \parencite[e.g.,][]{Chowkwanyun18,Rubens14,Lancet19,WongZ19,Flahault20,Khoury20,Morgenstern21,Slavkin20}. In this way it is similar to the predicted impact of ‘big data’ in public health, a good overview of which is provided by Dolley \parencite{Dolley18}. Zandi and colleagues \parencite*{Zandi19} capture the promissory potential in their call for papers on ethical challenges of AI in public health:
\begin{quote}
    These technologies promise great benefits to the practice of medicine and to the health of populations. This is especially true in epidemiology and the tracking of outbreaks of infectious diseases, behavioural science, precision medicine and the modelling and treatment of rare and/or chronic diseases.
\end{quote}
When combined with big data, AI approaches are expected to offer new opportunities to revolutionise epidemiology \parencite{Brockmann20} and to enable measurement of the impact of upstream determinants of health over the lifecourse \parencite{Krieger17,Meagher17,StephensonForthcoming}. Importantly, this would be a way of quantifying and revealing the “structured chances” that “drive population distributions of health, disease, and well-being” \parencite{Krieger12}. Recent examples of this have been the use of machine learning to quantify the relationship between social determinants and unmet dental care needs \parencite{Hung20}, prospective risk stratification for health plan payments \parencite{Irvin20}, and prediction of COVID-19 outcomes based on sociodemographics \parencite{Makridis21}. This opportunity arises because deep learning in particular offers novel capabilities to deal with complex, high-dimensional data and relatively small sample sizes \parencite{Ching18,LeCun15} – so called 'wide' data. The most aspirational accounts predict these new approaches will be able to facilitate action on social and environmental determinants of health, and thereby reduce health disparities \parencite{Katsis17,Meagher17,StephensonForthcoming,Weeramanthri18,Galea20}.

However, despite this potential being recognised, public health has been comparatively slow to broadly adopt AI in practice \parencite{Lavigne19,Panch19AIOppRisks,Sadilek18,Lancet19}. Predicting and tracking infectious disease outbreaks was an emerging application area for AI, and this has been greatly accelerated by the COVID-19 pandemic, with an AI epidemiological tool claiming to have been the first to sound warnings about the outbreak in Wuhan \parencite{Niiler20}. Key applications of AI in public health are:
\begin{itemize}
    \item population health surveillance and disease detection \parencite[e.g.,][]{BenAmmar18,Lake19,Muller19,Khadidos20,Perlman17,Sadilek18,Salathe16,Subramani18,Thorpe17,Xiong18,Baclic20,Munadi20}, including COVID-19 \parencite[e.g.,][]{Banerjee20}; 
    \item predicting and tracking infectious disease outbreaks \parencite[e.g.,][]{Bates17,Lim17,Park18,WongZ19}, including COVID-19 \parencite[e.g.,][]{Khadidos20}; 
    \item primary and secondary prevention of disease \parencite[e.g.,][]{Barrett13,Chatelan18,Contreras18,Meagher17,Potash15};
    \item environmental health \parencite[e.g.,][]{Dong18,Gonzalez-Jimenez18,KamelBoulos19,Li17DeepAirPollution,Sincak14,Weichenthal19};
    \item disease screening \parencite[e.g.,][]{McKinney20,Tran18,WongT16,Wu19DeepNeuralNetworks,Freeman21}; and
    \item risk factor intervention and treatment adherence \parencite[e.g.,][]{Deb18,Huang18,Labovitz17,Thompson19}.
\end{itemize}
An illustrative example is McKinney and colleagues \parencite*{McKinney20} demonstrating material reductions in the rates of false positives and false negatives using an AI system for breast cancer screening, highlighting AI's potential to improve the efficacy and cost-effectiveness of breast cancer screening programs, acknowledging however that robust evidence is lacking \parencite{Freeman21}. There have also been a number of demonstrations of the capability of AI to achieve accurate risk prediction \parencite[e.g.,][]{Attiga18,Chandir18,Harrington18,Khourdifi18,Kwon18,Miotto16,Nadkarni19,Pergialiotis18,Prelot18,Rajiwall17,Rajiwall18,WalshC17Predicting,Wiens16,Yu16}. Weng and colleagues \parencite*{Weng17} exemplified this capability by demonstrating that machine learning approaches could use routine clinical data to significantly improve the accuracy of cardiovascular risk prediction, compared to an established algorithm.

Risk prediction algorithms are also increasingly being used as a first line of automated triage in advance of primary care appointments \parencite[p. 23]{Derrington17}. For example, UK-based company Babylon Health\footnote{\url{https://www.babylonhealth.com/}} has a partnership with the UK’s National Health Service (NHS) called ‘GP at hand’ to provide online general practice consultations, with over 35,000 registered members as at January 2019 \parencite{Babylon19Presentation}. Babylon Health uses a digital symptom checker underpinned by AI to triage patients – this is an example of an automated decision system. Although concerns have been raised about the safety of digital symptom checkers \parencite{Fraser18}, in mid-2019 Babylon Health was able to raise an additional US\$550m in investment capital in order to enable the company to expand into the United States and develop the capability of its AI to diagnose more serious conditions \parencite{Lunden19}. Another application is automated prescribing of contraceptives. A small-scale study published in September 2019 in the \textit{New England Journal of Medicine} evaluated the safety of telecontraception, which involves the automated prescribing of contraceptives with or without clinicians in the loop. The study found that telecontraception may increase the accessibility of contraception, and also promote better adherence to treatment guidelines compared to in-person clinics \parencite{Jain19}.

AI-based risk stratification is also being used to enable automated, risk-adjusted, per capita funding allocation for health services and primary care. In this application the amount of money allocated to people for primary care services for a period of time is assigned based on their health status and algorithmic predictions of risk. For example, a commercial algorithm is used by a number of Accountable Care Organisations in America to make healthcare resourcing decisions for over 70 million people \parencite{Obermeyer19,Obermeyer19FAT}. As another example, the Australian Government recently trialled risk-adjusted funding for primary care through the Health Care Homes initiative.\footnote{\url{http://www.health.gov.au/internet/main/publishing.nsf/Content/health-care-homes}} The amount of funding provided to participating general practitioners to coordinate the care of individual patients will be decided using a predictive risk algorithm. The algorithm – developed by the CSIRO\footnote{\url{https://www.csiro.au/}} – factors in more than 50 variables, including demographics, a proxy for social determinants (the Australian Bureau of Statistics’ SEIFA indices for social and economic status\footnote{See Australian Bureau of Statistics \parencite*{ABS18SEIFA}}), physiology, medicines, conditions, pathology results, and lifestyle factors \parencite{Hibbert18}. 

In summary, accounts in literature and the media reveal a tremendous expectation on AI to transform medical and public health practices. There have been prominent demonstrations of successful narrow AI capability in medical and public health applications – particularly in diagnostic decision making, risk prediction and disease surveillance. These demonstrations reinforce the hype and expectation surrounding AI, and stimulate its rapidly expanding adoption in medicine and public health. This rapid adoption reinforces the need to carefully consider the longer-term hazards specific to public health.

\subsection{Emerging issues}
As the adoption of narrow AI and automated decision making in sensitive domains expands, this review has found that significant evidence of emerging issues has been gathered, including in health applications. Indeed, Whittaker and colleagues \parencite[p. 42]{Whittaker18Report} contend that the harms and biases in AI systems are now beyond question. “That debate has been settled,” they write, “the evidence has mounted beyond doubt”. They point to a growing consensus – citing a string of high-profile examples – that AI systems are perpetuating and amplifying inequities \parencite{AINow18YearReview,Whittaker18Report}. This review will now focus on three key issues which have emerged in the analysis phase: 1) bias, 2) opacity and incontestability, and 3) erosion of privacy – as these appear to be materialising in medical and public health applications of AI, and also because of the potential for these issues to give rise to mechanisms by which existing health inequities are  entrenched and amplified, engendering potential risk for population health equity.

\subsubsection{Bias}
Of the issues that have emerged in the application of AI and automated decision making, bias is perhaps the most prominent in the literature reviewed. Defining bias is difficult because the term has specific meanings in fields such as statistics, epidemiology, and psychology, and these are often confusingly contradictory \parencite{Campolo17}. Whittaker and colleagues \parencite*{Whittaker18Report} distinguish between two types of bias arising from automated decision systems: allocative – where resources or opportunities are unfairly distributed; and representational – where harmful stereotypes and categorisations are reproduced and amplified.

The hope that AI will assist to \textit{overcome} biases in human decision making \parencite[e.g.,][]{Baur17,Lavanchy18,Luckin17} has been used as a justification for the use of automated decision systems \parencite[e.g.,][]{Lecher18}. However, there have been glaring examples of racial, gender and socioeconomic biases evident in AI and automated decision making used in a number of sensitive domains, including: 
\begin{itemize}
    \item criminal justice \parencite{Angwin16,EPIC17,Eubanks17,Lapowsky18,Lum16};
    \item policing \parencite{BennettMoses18,RichardsonForthcoming,Sentas17,Stanley18,Vogt18Part1,Vogt18Part2};
    \item hiring practice \parencite{Dastin18,Mann16};
    \item university admissions \parencite{Schwartz19};
    \item online advertising \parencite{Bolukbasi16,Campolo17,Lambrecht18};
    \item education \parencite{AINow18Litigating,Campolo17,Feathers19,Madnani17,Lamont21};
    \item immigration \parencite{Whittaker18Report}; and
    \item facial recognition \parencite{Buolamwini18,Concerned19Letter,Garvie16,Lohr18,Singer18,Vincent19}.
\end{itemize}
There is also emerging evidence of the harmful impact of biases in the context of algorithmic censorship \parencite{Binns17,Cobbe19}. For example, there is racial bias in how hate speech is moderated \parencite{Sap19}, gender bias in how nudity is censored on Instagram \parencite{Cook19,Toor16}, and censorship of marginalised communities through overly-restrictive automated filtering of LGBTQ content on YouTube, Tumblr and Twitter \parencite{Allen18}.

Generally, algorithmic biases can arise in two main ways: in the upfront design (specification) of an algorithm, and in the data that are used to train algorithms, for example by being unrepresentative, or encoding existing systemic biases \parencite{Ankeny17,Crawford16Report,DawsonN18,McGoey17}. Bughin and colleagues \parencite*[p.37]{Bughin17} explain how bias can be caused by data: “Since the real world is racist, sexist, and biased in many other ways, real-world data that feeds algorithms will also have these features—and when AI algorithms learn from biased training data, they internalize the biases, exacerbating those problems.” As bias can arise unintentionally from data used to train the algorithms, it can be very difficult to detect and measure \parencite{Barocas16,Campolo17,Lecher18,Reisman18AlgoImpactAssess}. 

Algorithmic bias – especially undetected bias – can lead to inaccurate and inappropriate generalisation \parencite{Brooks17Machine,Khoury14,Maddox19,Muller19}. Generalisation is a key issue in machine learning theory and practice \parencite{Mackenzie15}. The general rigidity and brittleness of machine learning models means that models built for a specific purpose cannot be readily transferred to other applications, nor are they robust to changes over time \parencite{Brooks17Machine}. Barocas and Selbst \parencite*{Barocas16} make the crucial point that inappropriate generalisation is typically a result of careless reliance on “statistically sound inferences that are nevertheless inaccurate” (p. 688) – rather than deliberate prejudice. Again, that the disparate impact is inadvertent, makes it wickedly difficult to detect. And moreover, inappropriate generalisation can have a performative\footnote{Performativity, in this context, is the act of making a prediction having the effect of contributing to the predicted outcome coming into being.} impact \parencite{DawsonD19,Mackenzie15}, where inaccurate predictions actively contribute to \textit{produce} discriminatory outcomes. This is especially evident in criminal justice and predictive policing implementations of automated decision systems \parencite{Angwin16,BennettMoses18,Lum16,RichardsonForthcoming,Stanley18}. Inaccurate generalisation also stems from AI’s inherent reliance on data, and the axiomatic tension between over-fitting to past data and predictive accuracy. Writing for \textit{Computerworld}, George Nott quotes Genevieve Bell: "Humans can sometimes fear their choices are being "prescribed by their past" by these algorithms, which by their nature work on retrospective data" \parencite{Nott17}. The reliance on past data is a key reason why there is a risk that automated decision systems will perpetuate inequities, particularly where the systems rely on data that either reflects past systemic inequalities, or does not adequately encode social and environmental determinants \parencite{Chowkwanyun18}.

As with other high-stakes domains, bias has been called out as a key issue that will need to be addressed before AI can be trusted and more widely adopted in health \parencite{Campolo17,Challen19,Beam20,Brault20}. Specific to the health domain, numerous scholars have highlighted the lack of diversity, inclusiveness, and representativeness in health datasets \parencite[e.g.,][]{Barocas16,Campolo17,Dolley18,Lavigne19,LePage19,Meagher17,Panch19InconvientTruth,Prainsack19,Walsh19ACOLA,Whittaker18Report,Kiang21,Lee20,Pham21,Ti20}. And as previously noted, the use of biased data is known to reproduce and amplify discrimination and injustice \parencite{Barocas16,Crawford16Report,Jasanoff17}. For instance, Straw and Callison-Burch \parencite{Straw20} demonstrated the existence of significant biases in natural language processing models used in psychiatry, and identified the risk that these biases may widen health inequalities. 

In public health too, it is well-recognised that skewed and unrepresentative data can bias the results of traditional epidemiological and population health analyses such as disease surveillance, leading to inaccurate estimates and inference for diverse populations \parencite{Bates17,Krieger12,Thorpe17}. Exemplifying how data quality can affect automated decision systems, flawed data was blamed for the failures of Idaho’s automated decision system to equitably allocate home care funding \parencite{Stanley17}. And in a study that has striking similarities to \textit{ProPublica’s} revelatory investigative reporting into racially biased crime risk prediction \parencite{Angwin16}, Obermeyer and colleagues \parencite*{Obermeyer19} detected significant racial bias in a commercial algorithm used by Accountable Care Organisations in America and applied to an estimated 200 million people each year. Their analysis revealed that White patients were given the same risk score as Black patients who were considerable sicker, inadvertently leading to Black patients having unequal access to care. The authors estimated that resolving this disparity would have more than doubled the proportion of Black patients receiving additional assistance (from 17.7\% to 46.5\%). The paucity of environmental and social exposure data has also been identified \parencite[e.g.,][]{AIHW18,Derrington17}, however the potential for this to lead to biases in narrow AI and automated decision systems needs to be further explored \parencite{Pham21}.

\subsubsection{Opacity and incontestability}
Another key issue is the opacity of artificial intelligence, and the ensuing incontestability of automated decisions. Algorithms and AI are opaque and invisible processes, often characterised as ‘black boxes’ \parencite{DawsonN18,Knight17,Pasquale15,Rahwan19,Salathe18}. Once an AI algorithm has been trained – particularly one based on deep learning – it is not clear how it is making decisions \parencite{Knight17,Waldrop19}. The lack of explainability of AI when applied in healthcare has been identified as a threat to the core ethical values of medicine \parencite{Amann20}. Research into public perceptions reveal confusion amongst the general public about the inner workings of algorithms, and wariness about inscrutable algorithmic processes that have delegated responsibility for high-stakes decisions \parencite{SmithA19,Araujo20,McCradden20}. 

A consequence of this opacity is the difficulty of questioning and contesting automated decisions. Whittaker and colleagues at the AI Now Institute observe that when automated decision systems make errors, “the ability to question, contest, and remedy these is often difficult or impossible” \parencite{Whittaker18Report}. This is exemplified in the United States criminal justice system, where “Defendants rarely have an opportunity to challenge their [algorithmic] assessments” \parencite{Angwin16}. Furthermore, the ability of humans to intervene, override or even explain decisions is severely limited, rendering frontline workers disempowered intermediaries \parencite{Whittaker18Report}. Early reports suggest that issues of incontestability have emerged in the use of automated decision systems in health. This powerlessness is evident in Colin Lecher’s article for \textit{The Verge} \parencite*{Lecher18} about the case of a women with cerebral palsy who had her health services funding cut in half by an automated algorithmic decision. When an attorney began to investigate complaints about the algorithm, he found: “No one seemed able to answer basic questions about the process. The nurses said, ‘It’s not me; it’s the computer’.”

For people who are the subjects of automated decisions, there is even more of a sense of powerlessness. Regarding the American Civil Liberties Union (ACLU) legal challenge to Idaho’s use of an algorithmic decision system to allocate home care funding \parencite[see also][]{Stanley17}, Lecher \parencite*{Lecher18} writes: “Most importantly, when Idaho’s system went haywire, it was impossible for the average person to understand or challenge”. Incontestability can therefore lead to aggregation of power, limited opportunities for redress, and an unwillingness and inability of vulnerable people to contest their own treatment, thereby perpetuating and exacerbating existing inequities and discriminatory dynamics \parencite{Crawford16Report,Eubanks17}.

The tendency to blindly trust complex statistical methodologies both fortifies the inscrutability of automated decisions, and also intensifies the performativity of prediction. In the field of public health, concern has been raised about overconfidence in big data and complex statistical techniques \parencite{Chiolero18,Lavigne19}. Salathé \parencite*{Salathe16} refers to this as “big-data hubris”. Similarly, Krieger \parencite*{Krieger17}, quoting prescient statistician Lancelot Hogben, cautions against hiding “behind an impressive façade of flawless algebra”. Artificial intelligence systems, because they are considered ‘intelligent technology’, are particularly prone to going uncontested \parencite[pp. 6-7]{Crawford16Report}. Underlying this misplaced trust is a reductionist belief in the neutrality of data – a belief in data being beyond reproach. Sheila Jasanoff \parencite*{Jasanoff17} captures this eloquently: 
\begin{quote}
    ...in modernity, information, along with its close correlate data, has been taken for granted as a set of truth claims about the way the world is. Information, as conventionally understood, quite simply is what \textit{is}: it consists of valid observations about what the world is like. Data represents a specific form of information, a compilation of particular types of facts designed to shed light on identifiable issues or problems. As representations of reality, both public information and public data were seen until recently as lying to some extent outside the normal domains of political inquiry. (p. 5, emphasis in the original)
\end{quote}
But data are not neutral \parencite{Barrowman18,boyd12,Zook17}. Krieger \parencite{Krieger12} for one cautions against treating populations as statistical entities comprised only of data, calling out the “uncritical approach to presenting and interpreting population data, premised on the dominant assumption that population rates are statistical phenomena driven by innate individual characteristics.”

What follows from overconfidence in complex statistical techniques and belief in the neutrality of data is a misplaced trust in the ability of automated decision systems to make correct, unbiased decisions. Virginia Eubanks is quoted by Lecher \parencite*{Lecher18} as saying that “there is a “natural trust” that computer-based systems will produce unbiased, neutral results.” Likewise, Campolo and colleagues \parencite*{Campolo17}, citing Sandra Mayson’s work on algorithmic risk assessment in setting bail, point out the potential of risk assessment to “legitimize and entrench” problematic reliance on statistical correlation, and to "[lend such assessments] the aura of scientific reliability.” And similarly, Barocas and Selbst \parencite*{Barocas16} identify the “imprimatur of impartiality” conferred on the decisions taken by algorithmic systems. This is important because it gives rise to a false confidence in the superiority of automated decisions. 

Through their complexity, invisibility, and inhumanity, algorithmic decisions are achieving incontestability. And thus, when predictions are made, they verge on acts of creation, of magic \parencite{Chun11}. Will Knight wrote in 2017: “As the technology advances, we might soon cross some threshold beyond which using AI requires a leap of faith.” Considering how “indecipherable” algorithmic systems can be \parencite{Rahwan19}, and how they can be “beyond the understanding even of the people using them” \parencite{Lecher18}, it should therefore not be surprising that the use of automated decision systems creates legal uncertainty by challenging conventional models of autonomy \parencite{Arnold21} and  creating a growing “accountability gap" \parencite{Whittaker18Report}. This is perpetuated by trade secrecy and intellectual property provisions that enable proprietary systems to be shielded from scrutiny, even in the face of legal challenge \parencite{Angwin16,Campolo17,DawsonD19,Muller19,Obermeyer19,Rahwan19,Salathe18,Whittaker18Report}. The research of the AI Now Institute has uncovered “black boxes stacked on black boxes: not just at the algorithmic level, but also trade secret law and untraceable supply chains.” \parencite{AINow18YearReview} In this way, trade secrecy reinforces the incontestability of automated decisions \parencite{Whittaker18Report}, which heightens the risk that existing biases and disparities are perpetuated and amplified through disproportionately affecting vulnerable and at-risk populations.

\subsubsection{Privacy erosion}
The use of artificial intelligence is also having a significant impact on human rights to privacy, freedom of expression, and access to information \parencite{OVIC18,PrivacyIntl16,PrivacyIntl18,Santow18}. While privacy issues are not limited to AI, three dynamics eroding privacy: re-identification risk, intrusive data extraction and capitalisation, and invasive surveillance – stand out from the literature in relation to AI. Firstly, many scholars and institutions have highlighted the risk of re-identification and compromising individual privacy which arise from big data analytics and AI \parencite[e.g.,][]{DawsonD19,Dolley18,Mittelstadt2018,Ohm10,Rocher19,Zook17}. Secondly, expanding use of AI in surveillance – for example use of facial recognition in policing \parencite{DawsonD19}, and monitoring of employees’ emotional state for performance evaluation and retention decisions \parencite{Campolo17} – is eroding privacy and amplifying discriminatory dynamics \parencite{DawsonD19,Whittaker18Report,Zuboff19}. And thirdly, as a result of the recognition of and the use of AI to exploit the economic value of personal data \parencite{Leonelli19,WEF11}, systems for data extraction, ‘datafication’ and capitalisation are becoming increasingly intrusive and pervasive \parencite{Jasanoff17,Newell15,Parry17,Sadowski19,Schneier19Data,West17,Zuboff19}. This intrusiveness and exploitation has resulted in growing community wariness of data sharing, and erosion of social licence for use of individual data \parencite{IPSOS16,Richmond19,SmithA19}.

Reports about erosion of privacy are also already prevalent in health applications of narrow AI. Like other domains, the value of personal health data has long been recognised \parencite{Leonelli19,McMorrow14,WEF11,Thapa21,Zeitoun20}. The extraction of data will be driven more and more by commercialisation and productisation of data as a tradable asset \parencite{Newell15,Parry17,Sadowski19,West17,Zuboff19}. In health, this has seen the emergence of specialist data brokers, such as Explorys\footnote{\url{https://www.ibm.com/watson/health/explorys/}}, which was purchased by IBM in 2015. An example brokerage is Memorial Sloan Kettering (MSK) Cancer Center entering into a licencing agreement with AI start-up Paige.AI\footnote{\url{https://paige.ai/}} to “grant exclusive access to MSK’s intellectual property in computational pathology, including access to MSK’s 25 million pathology slides.” \parencite{MedTech18}. However, the lure of using personal health data in AI and big data applications is driving an erosion of rights to privacy, confidentiality, and data ownership \parencite{Balthazar18,Bogle19,Sharon18,Brockmann20}. For example, there are increasing privacy concerns expressed in the media and literature about health apps’ lack of transparency around data sharing and use \parencite{Grundy19}. Another example is data sharing between the UK National Health Service and Google DeepMind being considered a betrayal of public trust \parencite{Hern18,Lomas18,Revell17}. Google has also faced media criticism for gathering personal health information on millions of people in the United States as part of 'Project Nightingale' \parencite{Anonymous19Guardian,Copeland19,Fussell19}, as has Memorial Sloan Kettering health service for its data sharing arrangement with Paige.AI \parencite{Ornstein18}. Throughout the COVID-19 pandemic there has also been widespread debate regarding the impact on privacy of using digital tracking tools including AI \parencite[e.g.,][]{Smidt21}. Campolo and colleagues \parencite*{Campolo17} point out that in domains like health, because of AI’s reliance on large amounts of data, the privacy rights of vulnerable populations are particularly at risk due to lack of informed consent, inability to opt out, and poor due process mechanisms. In this way, erosion of privacy and commodification of individuals’ data will likely disproportionately affect vulnerable populations.

\section{Discussion}
Increasingly frequent and prominent demonstrations of narrow AI capability in medicine and public health are stimulating a rapid expansion in their adoption, reinforced by economic drivers and accelerated by the COVID-19 pandemic. The examples given in section 3.1 of automated decision systems that allocate health services funding illustrate how decisions which are automatically made based on the results of an algorithm can be consequential for population health. In the example of the Health Care Homes program in Australia, through its scale, decisions have meaningful consequences for population health and wellbeing, for the livelihood of general practitioners, and for the sustainability of the primary care tier of the Australian health system. Malfunctioning of these systems would therefore be expected to adversely affect population health equity.

Meanwhile, it is clear from the results of this review that there is significant evidence of and concern about issues that have emerged in the use of narrow AI and automated decision making in sensitive domains where AI is already widely utilised. As the use of narrow AI-based automated decision systems and algorithmic prediction in health continues to expand, it is probable that the same issues which have demonstrably emerged in other sensitive domains will also manifest widely in medicine and public health applications. Indeed, there is emerging evidence of this happening already – that it is not more widespread is perhaps due to the comparatively slower adoption of AI in public health.

There are two key reasons why it is highly probable that issues such as bias, incontestability and erosion of privacy will manifest widely in health. Firstly, as outlined in the results section, early reports of issues associated with the use of automated decision systems and predictive algorithms in health have already surfaced, and it is reasonable to expect this trend to continue. Secondly, the same circumstances and drivers which have compelled adoption of automated decision systems and given rise to issues in other high-stakes domains, also exist in medicine and public health – indicating the adoption trajectory and consequences are likely to be similar. Key amongst these drivers are cost and capacity pressures facing health services, and the commercial imperative to use AI to capitalise on growing health data assets.

The same imperatives to constrain costs and capitalise on data exist in other sensitive domains – such as education, policing, criminal justice, and immigration – and this has incentivised the adoption of automated decision systems by public agencies and corporations in those domains \parencite{Reisman18AlgoImpactAssess,Whittaker18Report}. A process of learning and emulation akin to policy transfer\footnote{Policy transfer can be defined as a process of learning and emulation by policy makers in which “knowledge about policies, administrative arrangements, institutions etc. in one time and/or place is used in the development of policies, administrative arrangements and institutions in another time and/or place.” \parencite{Dolowitz96}} will likely ensure the adoption trajectory of automated decision systems will be similar in the health domain. This emulation and diffusion of innovation occurs because jurisdictions and agencies “face common problems” and they look to other communities for lessons and solutions \parencite{Hadjiisky17}. The process is accelerated by a futures industry whose purpose is to market ideas and trade on promises and expectation. Hadjiisky and colleagues write:
\begin{quote}
    ...an entire global marketplace of ideas and recommendations on ‘best practices’ has emerged, including international organizations, commissions, donor groups, consultants, think tanks, institutes, networks, partnerships, and various gatherings of the great and the good such as Davos. They may not use the terminology of ‘policy transfer’ but that, in essence, is what they are debating and selling. (p. 2-3)
\end{quote}
Indeed, there is tremendous expectation heaped upon AI and precision health approaches to increase the efficiency of healthcare services and systems as a means of containing costs \parencite{Lecher18,Panch19InconvientTruth,Prainsack19,Whittaker18Report}. Dolley \parencite*{Dolley18} captures this in relation to public health:
\begin{quote}
    Precision public health is exciting. Today’s public health programs can achieve new levels of speed and accuracy not plausible a decade ago. Adding precision to many parts of public health engagement has led and will lead to tangible benefits. Precision can enable public health programs to maintain the same efficacy while decreasing costs, or hold costs constant while delivering better, smarter, faster, and different education, cures and interventions, saving lives. (p. 6)
\end{quote}
The drive to rapidly adopt AI in health is given urgency by the oft-cited pressures facing health systems around the world, including population ageing, workforce shortages, increased prevalence and incidence of noncommunicable diseases, and variability in service quality and clinical outcomes \parencite{ACSQHC18,Britnell19,CSIRO18,Lancet19}. These pressures are especially acute in low-income countries, where health resources are particularly scare \parencite{Lancet19}, contributing to the expectation that AI will benefit global health \parencite{Alami20,Mehta20,Schwalbe20}. Similarly, long-standing problems with current diagnostic approaches, such as invasiveness, cost, accessibility, and low precision, as well as the limitations of traditional analytic approaches, are driving interest in improved AI-enabled methods \parencite{Derrington17,Maddox19}. The COVID-19 pandemic has been a ‘perfect storm’ that has exacerbated these pressures, further accelerating the adoption of AI \parencite{Horgan20,IqbalS21,Peek20,Saniotis20,Scott20}.

The drive to adopt AI in health also follows closely on the heels of the imperative in medicine and public health to capitalise on big data. Much like AI more recently, 'big data' has commonly been expected to transform medicine and public health practice \parencite[e.g.,][]{Bates17,Choucair15,Dolley18,Khoury14,Mattick14,Salathe18,Thorpe17,Weeramanthri18,Williamson18}. AI – particularly deep learning – promises the ability to \textit{finally} exploit big, complex, noisy, highly-dimensional health data that health organisations have been accumulating \parencite{Katsis17,Lavigne19,Ogino19,Salathe18,Thorpe17,WongZ19}. For example, an editorial in \textit{The Lancet Public Health} \parencite*{Lancet19} states: “The ability of artificial intelligence and machine learning algorithms to analyse these multiple and rich data types at a scale not previously possible could bring a step change in public health and epidemiology.” There is a convergence between the accumulation of data \parencite{Sadowski19} and hyper-enthusiasm about AI. The drive to make use of and assetize data in health \parencite{Tarkkala19,Vezyridis21} will drive adoption of AI and automated decision systems. Typifying this drive is a call by the Chief Executive of a new United Kingdom National health Service agency, NHSX, to capitalise on big data and AI \parencite{SmithR19}. Underscoring the financial imperative pressuring health services and agencies to adopt AI, the economic opportunity has been identified not only by corporations, but by governments and public agencies \parencite[e.g. in Australia:][]{AlphaBeta18,CSIRO18,SenateSelectComm16}.

What the drive to address health system pressures and capitalise on data portends is that the adoption of AI systems will be substantiated on the basis of health service efficiency and productivity – chasing “commercial vistas of fabulous scale” \parencite{Kissinger18} – and not necessarily on improving population health or population health equity. Obermeyer and colleagues’ \parencite*{Obermeyer19} analysis already exhibits the perverse outcomes resulting from optimisation of AI prediction based on health service cost as a goal function, and not population health outcomes. Because of the likely focus on efficiency and headlong rush toward adoption, issues may well go overlooked or even ignored. 

So why does it matter that issues such as bias, incontestability, and privacy erosion manifest widely? It matters because these issues give rise to mechanisms by which existing social, economic and health inequities are perpetuated and amplified, and also potentially create new inequities. Bias does this by perpetuating and amplifying existing inequities where automated decision systems rely on data that unknowingly echo systemic inequities, or do not adequately take account of the social and environmental determinants of health which drive inequitable health disparities. The opacity and incontestability of automated decision systems disproportionately affect vulnerable populations through aggregation of power, limiting opportunities for redress, and the inability of vulnerable people to contest unjust decisions. Similarly, vulnerable populations are disproportionately at risk of privacy erosion and data exploitation through the use of AI. Together these issues have the potential to perpetuate and exacerbate existing inequities and discriminatory dynamics. This leads to a significant risk that use of automated decision systems in health will amplify and entrench existing population health inequities.

Examples of population health inequities that may be affected include the persistent life expectancy gap afflicting Aboriginal and Torres Strait Islander peoples in Australia \parencite{Holland18,Lowitja19}, and the stark socio-economic gradient in health outcomes evident in many countries \parencite{CoSD08,Marmot05,NRCandIOM2013}, through, for example, social and environmental determinants such as intergenerational trauma not being accounted for in algorithmic allocation of health resources. In addition, because of the strong influence of social determinants of health \parencite{CoSD08}, the use of automated decision systems in other domains such as welfare and immigration, will undoubtedly also have a downstream effect on population health equity \parencite[see for example][]{Medhora19}, either because these determinants are missing in the data, or the data lack diversity, inclusiveness, and representativeness.

Previous scholarly warnings regarding the risk that ‘precision’ health approaches have the potential to exacerbate health inequities lends credence to the existence of the risk to population health equity posed by narrow AI and automated decision systems. There have been strong warnings from within public health (albeit with little empirical evidence of their manifesting to date) that ‘precision medicine’, as well as emerging ‘precision public health’ and ‘precision prevention’ approaches (which employ narrow AI, automated decision systems and algorithmic risk prediction) have the potential to produce disparate impacts, amplify existing prejudices, and propagate health inequities \parencite{Jasanoff17,Khoury16,Meagher17,Panch19AIOppRisks,Prainsack17,Prainsack19,TaylorRobinson18,Lancet19}. Elucidating this, Lavigne and colleagues \parencite*{Lavigne19} write:
\begin{quote}
    ...particularly when applying these approaches to decision-making or predictions at a population level, attention must be paid to the potential for these approaches to produce health inequities, either through the use of biased data or through uneven access to the technology. Predictions and models based on non-representative or biased data can propagate underlying biases and exacerbate health inequities at a population level if sufficient care is not taken to mitigate these issues. (p. 176)
\end{quote}
Moreover, socioeconomic gradients in access to, as well as the means and resources to best utilise new precision health tools – for example genomic risk prediction – have the potential to widen inequalities further \parencite{Meagher17,Prainsack19,Vayena15}. The focus on individual risk factors promoted by precision approaches can also reinforce the notion of individual responsibility, prolonging the use of individualist, behaviourist interventions, which tend to entrench and exacerbate socioeconomic disparities in health outcomes \parencite{Baum14,Dolley18,Meagher17,Prainsack19,StephensonForthcoming,TaylorRobinson18}. The focus on individual risk factors also undermines the rationale and societal propensity to act on structural, upstream determinants of health inequities, thereby permitting inequities in population health to persist \parencite{Chowkwanyun18,Khoury16,Meagher17,Panch19AIOppRisks}. Meagher and colleagues \parencite*[][p. 11]{Meagher17} succinctly capture this idea in relation to genomic data: “genomic explanations for health disparities can distract and even exculpate society from taking responsibility for the structural determinants of those inequities, undermining the political momentum of those seeking justice”. This same rationale applies to the use of AI and automated decision systems.

\section{Conclusion}
Amid unprecedented global interest in artificial intelligence, there are tremendous expectations that AI will transform medicine and public health practice. And while it may go largely unnoticed, narrow AI and predictive algorithms are already ubiquitous; woven into the fabric of our daily lives. Decisions which are being made automatically about disease detection, diagnosis, treatment and funding allocation have significant consequences for individual and population health and wellbeing.

The evidence collated in this review makes it clear that issues have emerged in sensitive domains like criminal justice where narrow AI and automated decision systems are already in common use. As their use in health rapidly expands, it is probable that the same issues – bias, incontestability, and privacy erosion – will also manifest widely in medicine and public health applications. Reports of this happening are already appearing. Moreover, the combination of hype, the drive to adopt automated decision systems to address cost pressures – accelerated by the COVID-19 pandemic – and the commercial imperative to capitalise on health data assets, may conspire to obscure issues, as has occurred in other domains.

Crucially, bias, incontestability, and erosion of privacy give rise to mechanisms by which existing social, economic and health disparities are perpetuated and amplified by automated decision systems. Therefore, there is a significant risk that the use of automated decision systems in health will exacerbate existing population health inequities and potentially create new ones. Medical and public health interventions have obviously produced disparate outcomes in the past; what makes the risk with narrow AI and automated decision systems \textit{different} is the industrial scale and rapidity \parencite{Kissinger18} with which they can be applied to whole populations, combined with the incontestability of decisions. This means negative consequences can quickly escalate to affect whole populations.

While it is too soon to say whether the issues emerging in health applications of narrow AI and automated decision systems have led to worsened population health inequity, it is incumbent on health practitioners and policy makers to explore and be mindful of the potential implications of using automated decision systems, so as to ensure the use of AI promotes population health and equity. There is a need to design and implement automated decision systems with care (using equity impact assessments for instance \parencite[see for example][]{HarrisRoxas11}), monitor their impact over time (especially longer-term effects on population health), and take responsibility for responding to issues as they emerge – even if this is long after a system has first been introduced. To finish, Obermeyer and colleagues \parencite*{Obermeyer19} set a very positive example. After uncovering inadvertent racial bias in an automated decision system allocating health assistance funding, they approached the algorithm manufacturer, who was able to independently replicate the results to confirm the existence of bias. The researchers and the algorithm manufacturer are now collaborating on developing solutions to address this bias. This is a fine example to emulate.

\section{Acknowledgements}
The author gratefully acknowledges the generous and valuable contributions of Niamh Stephenson, Alison Ritter, and kylie valentine.

\printbibliography

@report{ACSQHC18,
   author = {ACSQHC},
   year = {2018},
   title = {The Third Australian Atlas of Healthcare Variation},
   institution = {Australian Commission on Safety and Quality in Health Care},
   loction = {Sydney},
   url = {https://www.safetyandquality.gov.au/atlas/the-third-australian-atlas-of-healthcare-variation-2018/},
   type = {Report}
}

@article{AINow18YearReview,
   author = {{AI Now Institute}},
   year = {2018},
   date = {25 Oct},
   title = {{AI} in 2018: A year in review},
   journaltitle = {Medium},
   url = {https://medium.com/@AINowInstitute/ai-in-2018-a-year-in-review-8b161ead2b4e},
   urldate = {2019-08-11},
   type = {Electronic Article}
}

@report{AINow18Litigating,
   author = {{AI Now Institute}},
   year = {2018},
   title = {Litigating Algorithms: Challenging Government Use of Algorithmic Decision Systems},
   institution = {{AI} Now Institute, New York University \& Center on Race, Inequality, and the Law \& Electronic Frontier Foundation},
   url = {https://ainowinstitute.org/reports.html},
   type = {Report}
}

@report{AIHW18,
   author = {AIHW},
   year = {2018},
   title = {Australia's Health 2018},
   number = {Australia’s health series no. 16. AUS 221},
   institution = {Australian Institute of Health and Welfare},
   loction = {Canberra},
   url = {https://www.aihw.gov.au/reports/australias-health/australias-health-2018},
   type = {Report}
}

@article{Allen18,
   author = {Allen, Samantha},
   year = {2018},
   date = {12 Oct},
   title = {Social Media Giants Have a Big LGBT Problem. Can They Solve It?},
   journaltitle = {The Daily Beast},
   url = {https://www.thedailybeast.com/social-media-giants-have-a-big-lgbt-problem-can-they-solve-it},
   urldate = {2019-09-20},
   type = {Electronic Article}
}

@report{AlphaBeta18,
   author = {{AlphaBeta Advisors}},
   year = {2018},
   title = {Digital Innovation: Australia’s \$315b Opportunity},
   institution = {CSIRO Data61},
   url = {https://www.data61.csiro.au/en/Our-Work/Future-Cities/Planning-sustainable-infrastructure/Digital-Innovation},
   type = {Report}
}

@article{Amoore09,
   author = {Amoore, Louise},
   year = {2009},
   title = {Algorithmic War: Everyday Geographies of the War on Terror},
   journaltitle = {Antipode},
   volume = {41},
   number = {1},
   pages = {49-69},
   DOI = {10.1111/j.1467-8330.2008.00655.x},
   type = {Journal Article}
}

@article{Angwin16,
   author = {Angwin, Julia and Larson, Jeff and Mattu, Surya and Kirchner, Lauren},
   year = {2016},
   date = {23 May},
   title = {Machine Bias},
   journaltitle = {ProPublica},
   publisher = {ProPublica},
   url = {https://www.propublica.org/article/machine-bias-risk-assessments-in-criminal-sentencing},
   urldate = {2018-10-14},
   type = {Electronic Article}
}

@article{Ankeny17,
   author = {Ankeny, Rachel A.},
   year = {2017},
   title = {Bringing Data Out of the Shadows},
   journaltitle = {Science, Technology, \& Human Values},
   volume = {42},
   number = {2},
   pages = {306-310},
   DOI = {10.1177/0162243916689138},
   type = {Journal Article}
}

@article{Anonymous19Guardian,
   author = {Anonymous},
   year = {2019},
   date = {15 Nov 2019},
   title = {I'm the Google whistleblower. The medical data of millions of Americans is at risk},
   journaltitle = {The Guardian},
   url = {https://www.theguardian.com/commentisfree/2019/nov/14/im-the-google-whistleblower-the-medical-data-of-millions-of-americans-is-at-risk},
   urldate = {2019-11-21},
   type = {Electronic Article}
}

@inproceedings{Attiga18,
   author = {Attiga, Y. and Chen, S. and LaGue, J. and Ovalle, A. and Stott, N. and Brander, T. and Khaled, A. and Tyagi, G. and Francis-Lyon, P.},
   year = {2018},
   title = {Applying Deep Learning to Public Health: Using Unbalanced Demographic Data to Predict Thyroid Disorder},
   booktitle = {IEEE 9th Annual Information Technology, Electronics and Mobile Communication Conference (IEMCON)},
   pages = {851-858},
   DOI = {10.1109/IEMCON.2018.8614888},
   type = {Conference Proceedings}
}

@online{ABS18SEIFA,
   author = {{Australian Bureau of Statistics}},
   year = {2018},
   date = {27 Mar},
   title = {Socio-Economic Indexes for Areas (SEIFA)},
   url = {http://www.abs.gov.au/websitedbs/censushome.nsf/home/seifa},
   urldate = {2019-02-21},
   type = {Web Page}
}

@article{Babylon19Presentation,
   author = {{Babylon Health}},
   year = {2019},
   date = {Jan},
   title = {Babylon GP at hand Progress to date [Presentation]},
   url = {https://assets-production.babylongpathand.co.uk/icons/GP-at-hand-Progress-to-date-Jan2019.pdf},
   urldate = {2019-03-29},
   type = {Electronic Article}
}

@article{Balthazar18,
   author = {Balthazar, P. and Harri, P. and Prater, A. and Safdar, N. M.},
   year = {2018},
   title = {Protecting Your Patients' Interests in the Era of Big Data, Artificial Intelligence, and Predictive Analytics},
   journaltitle = {J Am Coll Radiol},
   volume = {15},
   number = {3 Pt B},
   pages = {580-586},
   DOI = {10.1016/j.jacr.2017.11.035},
   type = {Journal Article}
}

@article{Barocas16,
   author = {Barocas, Solon and Selbst, Andrew D.},
   year = {2016},
   title = {Big Data's Disparate Impact},
   journaltitle = {California Law Review},
   volume = {671},
   DOI = {10.2139/ssrn.2477899},
   type = {Journal Article}
}

@article{Barrett13,
   author = {Barrett, Meredith A. and Humblet, Olivier and Hiatt, Robert A. and Adler, Nancy E.},
   year = {2013},
   title = {Big Data and Disease Prevention: From Quantified Self to Quantified Communities},
   journaltitle = {Big Data},
   volume = {1},
   number = {3},
   pages = {168-175},
   DOI = {10.1089/big.2013.0027},
   type = {Journal Article}
}

@article{Barrowman18,
   author = {Barrowman, Nick},
   year = {2018},
   title = {Why Data Is Never Raw},
   journaltitle = {The New Atlantis},
   number = {56},
   pages = {129-135},
   type = {Journal Article}
}

@article{Bates17,
   author = {Bates, M.},
   year = {2017},
   title = {Tracking Disease: Digital Epidemiology Offers New Promise in Predicting Outbreaks},
   journaltitle = {IEEE Pulse},
   volume = {8},
   number = {1},
   pages = {18-22},
   DOI = {10.1109/MPUL.2016.2627238},
   type = {Journal Article}
}

@article{Baum14,
   author = {Baum, F. and Fisher, M.},
   year = {2014},
   title = {Why behavioural health promotion endures despite its failure to reduce health inequities},
   journaltitle = {Sociol Health Illn},
   volume = {36},
   number = {2},
   pages = {213-25},
   DOI = {10.1111/1467-9566.12112},
   type = {Journal Article}
}

@article{Baur17,
   author = {Baur, Yvonne and Reid, Brenda and Hunt, Steve and Fitter, Fawn},
   year = {2017},
   date = {16 Jan},
   title = {How {AI} Can End Bias},
   journaltitle = {D!gitalist Magazine},
   url = {https://www.digitalistmag.com/executive-research/how-ai-can-end-bias},
   urldate = {2019-05-23},
   type = {Electronic Article}
}

@inproceedings{BenAmmar18,
   author = {Ben Ammar, R. and Ben Ayed, Y.},
   year = {2018},
   title = {Speech Processing for Early Alzheimer Disease Diagnosis: Machine Learning Based Approach},
   booktitle = {IEEE/ACS 15th International Conference on Computer Systems and Applications (AICCSA)},
   pages = {1-8},
   DOI = {10.1109/AICCSA.2018.8612831},
   type = {Conference Proceedings}
}

@article{BennettMoses18,
   author = {Bennett Moses, Lyria and Chan, Janet},
   year = {2018},
   title = {Algorithmic prediction in policing: assumptions, evaluation, and accountability},
   journaltitle = {Policing and Society},
   volume = {28},
   number = {7},
   pages = {806-822},
   DOI = {10.1080/10439463.2016.1253695},
   type = {Journal Article}
}

@report{Berg18,
   author = {Berg, Andrew and Buffie, Edward F. and Zanna, Luis-Felipe},
   year = {2018},
   title = {Should We Fear the Robot Revolution? (The Correct Answer is Yes)},
   number = {IMF Working Paper WP/18/116},
   institution = {International Monetary Fund},
   url = {https://www.imf.org/en/Publications/WP/Issues/2018/05/21/Should-We-Fear-the-Robot-Revolution-The-Correct-Answer-is-Yes-44923},
   type = {Report}
}

@inproceedings{Binns17,
   author = {Binns, Reuben and Veale, Michael and Van Kleek, Max and Shadbolt, Nigel},
   year = {2017},
   title = {Like Trainer, Like Bot? Inheritance of Bias in Algorithmic Content Moderation},
   booktitle = {9th International Conference on Social Informatics},
   loction = {Oxford UK},
   publisher = {Springer International Publishing},
   editor = {Ciampaglia, Giovanni Luca and Mashhadi, Afra and Yasseri, Taha},
   pages = {405-415},
   type = {Conference Proceedings}
}

@article{Bogle19,
   author = {Bogle, Ariel},
   year = {2019},
   date = {28 May},
   title = {Your health data was once between you and your doctor. But for how long?},
   journaltitle = {ABC News},
   url = {https://www.abc.net.au/news/science/2019-05-28/health-data-used-to-train-artificial-intelligence-privacy-risks/11102292?section=health},
   urldate = {2019-05-31},
   type = {Electronic Article}
}

@unpublished{Bolukbasi16,
   author = {Bolukbasi, Tolga and Chang, Kai-Wei and Zou, James and Saligrama, Venkatesh and Kalai, Adam},
   year = {2016},
   title = {Man is to Computer Programmer as Woman is to Homemaker? Debiasing Word Embeddings},
   howpublished = {arXiv},
   url = {https://arxiv.org/abs/1607.06520v1},
   type = {Preprint}
}

@book{Bostrom14,
   author = {Bostrom, Nick},
   year = {2014},
   title = {Superintelligence : paths, dangers, strategies},
   publisher = {Oxford University Press},
   location = {Oxford},
   edition = {1st},
   pages = {xvi, 328 pages},
   ISBN = {9780199678112
0199678111},
   type = {Book}
}

@article{boyd12,
   author = {{boyd}, {danah} and Crawford, Kate},
   year = {2012},
   title = {Critical questions for big data: Provocations for a cultural, technological, and scholarly phenomenon},
   journaltitle = {Information, Communication \& Society},
   volume = {15},
   number = {5},
   pages = {662-679},
   type = {Journal Article}
}

@article{Bridle18,
   author = {Bridle, James},
   year = {2018},
   date = {15 Jun},
   title = {Rise of the machines: has technology evolved beyond our control?},
   journaltitle = {The Guardian},
   url = {https://www.theguardian.com/books/2018/jun/15/rise-of-the-machines-has-technology-evolved-beyond-our-control},
   urldate = {2018-11-01},
   type = {Electronic Article}
}

@book{Britnell19,
   author = {Britnell, Mark},
   year = {2019},
   title = {Human: Solving the Global Workforce Crisis in Healthcare},
   publisher = {Oxford University Press},
   ISBN = {9780198836520},
   type = {Book}
}

@article{Brooks17Machine,
   author = {Brooks, Rodney A.},
   year = {2017},
   date = {28 Aug},
   title = {[FoR\&{AI}] Machine Learning Explained},
   journaltitle = {Robots, {AI}, and other stuff},
   url = {https://rodneybrooks.com/forai-machine-learning-explained/},
   urldate = {2018-10-11},
   note = {Blog},
   type = {Blog}
}

@article{Brundage15,
   author = {Brundage, Miles},
   year = {2015},
   title = {Taking superintelligence seriously: Superintelligence: Paths, dangers, strategies by Nick Bostrom (Oxford University Press, 2014)},
   journaltitle = {Futures},
   volume = {72},
   pages = {32-35},
   DOI = {https://doi.org/10.1016/j.futures.2015.07.009},
   type = {Journal Article}
}

@report{Bughin17,
   author = {Bughin, Jacques and Hazan, Eric and Ramaswamy, Sree and Chui, Michael and Allas, Tera and Dahlström, Peter and Henke, Nicolaus and Trench, Monica},
   year = {2017},
   title = {Artificial Intelligence : The Next Digital Frontier?},
   institution = {McKinsey Global Institute},
   url = {https://www.mckinsey.com/mgi/overview/2017-in-review/whats-next-in-digital-and-ai/artificial-intelligence-the-next-digital-frontier},
   type = {Report}
}

@inproceedings{Buolamwini18,
   author = {Buolamwini, Joy and Gebru, Timnit},
   year = {2018},
   title = {Gender Shades: Intersectional Accuracy Disparities in Commercial Gender Classification},
   booktitle = {1st Conference on Fairness, Accountability and Transparency},
   editor = {Sorelle, A. Friedler and Christo, Wilson},
   volume = {81},
   pages = {77--91},
   url = {http://proceedings.mlr.press},
   type = {Conference Proceedings}
}

@report{Campolo17,
   author = {Campolo, Alex and Sanfilippo, Madelyn and Whittaker, Meredith and Crawford, Kate},
   year = {2017},
   title = {{AI} Now 2017 Report},
   institution = {{AI} Now Institute, New York University},
   url = {https://ainowinstitute.org/AI\_Now\_2017\_Report.html},
   type = {Report}
}

@article{Challen19,
   author = {Challen, Robert and Denny, Joshua and Pitt, Martin and Gompels, Luke and Edwards, Tom and Tsaneva-Atanasova, Krasimira},
   year = {2019},
   title = {Artificial intelligence, bias and clinical safety},
   journaltitle = {BMJ Quality \& Safety},
   volume = {28},
   number = {3},
   pages = {231},
   DOI = {10.1136/bmjqs-2018-008370},
   type = {Journal Article}
}

@article{Chandir18,
   author = {Chandir, S. and Siddiqi, D. A. and Hussain, O. A. and Niazi, T. and Shah, M. T. and Dharma, V. K. and Habib, A. and Khan, A. J.},
   year = {2018},
   title = {Using predictive analytics to identify children at high risk of defaulting from a routine immunization program: Feasibility study},
   journaltitle = {Journal of Medical Internet Research},
   volume = {20},
   number = {9},
   DOI = {10.2196/publichealth.9681},
   type = {Journal Article}
}

@article{Chatelan18,
   author = {Chatelan, A. and Bochud, M. and Frohlich, K. L.},
   year = {2018},
   title = {Precision nutrition: hype or hope for public health interventions to reduce obesity?},
   journaltitle = {Int J Epidemiol},
   DOI = {10.1093/ije/dyy274},
   type = {Journal Article}
}

@article{Chilamkurthy18,
   author = {Chilamkurthy, Sasank and Ghosh, Rohit and Tanamala, Swetha and Biviji, Mustafa and Campeau, Norbert G. and Venugopal, Vasantha Kumar and Mahajan, Vidur and Rao, Pooja and Warier, Prashant},
   year = {2018},
   title = {Deep learning algorithms for detection of critical findings in head CT scans: a retrospective study},
   journaltitle = {The Lancet},
   volume = {392},
   number = {10162},
   pages = {2388-2396},
   DOI = {10.1016/S0140-6736(18)31645-3},
   type = {Journal Article}
}

@unpublished{Ching18,
   author = {Ching, Travers and Himmelstein, Daniel S. and Beaulieu-Jones, Brett K. and Kalinin, Alexandr A. and Do, Brian T. and Way, Gregory P. and Ferrero, Enrico and Agapow, Paul-Michael and Zietz, Michael and Hoffman, Michael M. and Xie, Wei and Rosen, Gail L. and Lengerich, Benjamin J. and Israeli, Johnny and Lanchantin, Jack and Woloszynek, Stephen and Carpenter, Anne E. and Shrikumar, Avanti and Xu, Jinbo and Cofer, Evan M. and Lavender, Christopher A. and Turaga, Srinivas C. and Alexandari, Amr M. and Lu, Zhiyong and Harris, David J. and DeCaprio, Dave and Qi, Yanjun and Kundaje, Anshul and Peng, Yifan and Wiley, Laura K. and Segler, Marwin H. S. and Boca, Simina M. and Swamidass, S. Joshua and Huang, Austin and Gitter, Anthony and Greene, Casey S.},
   year = {2018},
   title = {Opportunities And Obstacles For Deep Learning In Biology And Medicine},
   howpublished = {bioRxiv},
   DOI = {10.1101/142760},
   url = {http://biorxiv.org/content/early/2018/01/19/142760.abstract},
   type = {Preprint}
}

@article{Chiolero18,
   author = {Chiolero, Arnaud},
   year = {2018},
   title = {Data Are Not Enough-Hurray For Causality!},
   journaltitle = {American journal of public health},
   volume = {108},
   number = {5},
   pages = {622-622},
   DOI = {10.2105/AJPH.2018.304379},
   type = {Journal Article}
}

@article{Choucair15,
   author = {Choucair, Bechara and Bhatt, Jay D.},
   year = {2015},
   title = {A new era for population health: government, academia, and community moving upstream together},
   journaltitle = {American journal of public health},
   volume = {105 Suppl 2},
   number = {Suppl 2},
   pages = {S144-S144},
   DOI = {10.2105/AJPH.2015.302564},
   type = {Journal Article}
}

@article{Chowkwanyun18,
   author = {Chowkwanyun, Merlin and Bayer, Ronald and Galea, Sandro},
   year = {2018},
   title = {{“P}recision” Public Health — Between Novelty and Hype},
   journaltitle = {New England Journal of Medicine},
   volume = {379},
   number = {15},
   pages = {1398-1400},
   DOI = {10.1056/NEJMp1806634},
   type = {Journal Article}
}

@inbook{Chun11,
   author = {Chun, Wendy Hui Kyong},
   year = {2011},
   title = {Introduction: Software, a Supersensible Thing},
   booktitle = {Programmed Visions: Software and Memory},
   publisher = {The MIT Press},
   location = {Cambridge, Massachusetts \& London, England},
   pages = {1-11},
   url = {https://mitpress.mit.edu/books/programmed-visions},
   type = {Book Section}
}

@unpublished{Cobbe19,
   author = {Cobbe, Jennifer},
   year = {2019},
   title = {Algorithmic Censorship on Social Platforms: Power, Legitimacy, and Resistance},
   howpublished = {SSRN},
   DOI = {http://dx.doi.org/10.2139/ssrn.3437304},
   url = {https://ssrn.com/abstract=3437304},
   type = {Preprint}
}

@article{Coiera18,
   author = {Coiera, Enrico},
   year = {2018},
   title = {The fate of medicine in the time of {AI}},
   journaltitle = {The Lancet},
   volume = {392},
   number = {10162},
   pages = {2331-2332},
   DOI = {https://doi.org/10.1016/S0140-6736(18)31925-1},
   type = {Journal Article}
}

@report{CoSD08,
   author = {{Commission on Social Determinants of Health}},
   year = {2008},
   title = {Closing the gap in a generation: Health equity through action on the social determinants of health. Final Report of the Commission on Social Determinants of Health},
   institution = {World Health Organization},
   loction = {Geneva},
   url = {https://www.who.int/social\_determinants/thecommission/en/},
   type = {Report}
}

@article{Concerned19Letter,
   author = {{Concerned Researchers}},
   year = {2019},
   date = {26 Mar},
   title = {On Recent Research Auditing Commercial Facial Analysis Technology},
   journaltitle = {Medium},
   url = {https://medium.com/@bu64dcjrytwitb8/on-recent-research-auditing-commercial-facial-analysis-technology-19148bda1832},
   urldate = {2019-04-04},
   type = {Electronic Article}
}

@article{Contreras18,
   author = {Contreras, Ivan and Vehi, Josep},
   year = {2018},
   title = {Artificial Intelligence for Diabetes Management and Decision Support: Literature Review},
   journaltitle = {J Med Internet Res},
   volume = {20},
   number = {5},
   pages = {e10775},
   DOI = {10.2196/10775},
   type = {Journal Article}
}

@article{Cook19,
   author = {Cook, Jesselyn},
   year = {2019},
   date = {30 Apr},
   title = {Instagram’s Shadow Ban On Vaguely ‘Inappropriate’ Content Is Plainly Sexist},
   journaltitle = {HuffPost},
   url = {https://www.huffingtonpost.com.au/entry/instagram-shadow-ban-sexist\_n\_5cc72935e4b0537911491a4f},
   urldate = {2019-09-20},
   type = {Electronic Article}
}

@article{Copeland19,
   author = {Copeland, Rob},
   year = {2019},
   date = {11 Nov 2019},
   title = {Google’s ‘Project Nightingale’ Gathers Personal Health Data on Millions of Americans},
   journaltitle = {The Wall Street Journal},
   url = {https://www.wsj.com/articles/google-s-secret-project-nightingale-gathers-personal-health-data-on-millions-of-americans-11573496790},
   urldate = {2019-11-20},
   type = {Electronic Article}
}

@report{Crawford16Report,
   author = {Crawford, Kate and Whittaker, Meredith and Clare Elish, Madeleine and Barocas, Solon and Plasek, Aaron and Ferryman, Kadija},
   year = {2016},
   title = {The {AI} Now Report : The Social and Economic Implications of Artificial Intelligence Technologies in the Near-Term},
   institution = {{AI} Now Institute, New York University},
   url = {https://ainowinstitute.org/AI\_Now\_2016\_Report.html},
   type = {Report}
}

@report{CSIRO18,
   author = {CSIRO},
   year = {2018},
   title = {Future of Health : Shifting Australia’s focus from illness treatment to health and wellbeing management},
   institution = {CSIRO},
   url = {https://www.csiro.au/en/Do-business/Futures/Reports/Future-of-Health},
   type = {Report}
}

@article{Darcy16,
   author = {Darcy, A. M. and Louie, A. K. and Roberts, L. W.},
   year = {2016},
   title = {Machine Learning and the Profession of Medicine},
   journaltitle = {JAMA},
   volume = {315},
   number = {6},
   pages = {551-2},
   DOI = {10.1001/jama.2015.18421},
   type = {Journal Article}
}

@article{Dastin18,
   author = {Dastin, Jeffrey},
   year = {2018},
   date = {10 Oct},
   title = {Amazon scraps secret {AI} recruiting tool that showed bias against women},
   journaltitle = {Reuters},
   url = {https://www.reuters.com/article/us-amazon-com-jobs-automation-insight/amazon-scraps-secret-ai-recruiting-tool-that-showed-bias-against-women-idUSKCN1MK08G},
   urldate = {2018-11-16},
   type = {Electronic Article}
}

@report{DawsonD19,
   author = {Dawson, D and Schleiger, E and Horton, J  and McLaughlin, J and Robinson, C and Quezada, G and Scowcroft, J and Hajkowicz, S},
   year = {2019},
   title = {Artificial Intelligence: Australia’s Ethics Framework A Discussion Paper},
   institution = {CSIRO Data61},
   url = {https://www.data61.csiro.au/en/Our-Work/{AI}-Framework},
   type = {Report}
}

@report{DawsonN18,
   author = {Dawson, Nik},
   year = {2018},
   title = {{AI} Policy White Paper},
   institution = {Bits \& Atoms [website]},
   url = {https://bitsandatoms.co/ai-policy-white-paper-a-primer/},
   type = {Report}
}

@inproceedings{Deb18,
   author = {Deb, A. and Majmundar, A. and Seo, S. and Matsui, A. and Tandon, R. and Yan, S. and Allem, J. and Ferrara, E.},
   year = {2018},
   title = {Social Bots for Online Public Health Interventions},
   booktitle = {IEEE/ACM International Conference on Advances in Social Networks Analysis and Mining (ASONAM)},
   pages = {1-4},
   DOI = {10.1109/ASONAM.2018.8508382},
   type = {Conference Proceedings}
}

@report{Derrington17,
   author = {Derrington, Dolores},
   year = {2017},
   title = {Artificial Intelligence for Health and Health Care},
   number = {JSR-17-Task-002},
   institution = {{JASON} The MITRE Corporation},
   url = {https://fas.org/irp/agency/dod/jason/ai-health.pdf},
   type = {Report}
}

@article{Dolley18,
   author = {Dolley, Shawn},
   year = {2018},
   title = {Big Data’s Role in Precision Public Health},
   journaltitle = {Frontiers in Public Health},
   volume = {6},
   number = {68},
   DOI = {10.3389/fpubh.2018.00068},
   type = {Journal Article}
}

@article{Dolowitz96,
   author = {Dolowitz, David and Marsh, David},
   year = {1996},
   title = {Who Learns What from Whom: a Review of the Policy Transfer Literature},
   journaltitle = {Political Studies},
   volume = {44},
   number = {2},
   pages = {343-357},
   DOI = {10.1111/j.1467-9248.1996.tb00334.x},
   type = {Journal Article}
}

@inproceedings{Dong18,
   author = {Dong, S. and Feric, Z. and Li, X. and Rahman, S. M. and Li, G. and Wu, C. and Gu, A. Z. and Dy, J. and Kaeli, D. and Meeker, J. and Padilla, I. Y. and Cordero, J. and Vega, C. V. and Rosario, Z. and Alshawabkeh, A.},
   year = {2018},
   title = {A Hybrid Approach to Identifying Key Factors in Environmental Health Studies},
   booktitle = {IEEE International Conference on Big Data (Big Data)},
   pages = {2855-2862},
   DOI = {10.1109/BigData.2018.8622049},
   type = {Conference Proceedings}
}

@report{EPIC17,
   author = {EPIC},
   year = {2017},
   title = {Algorithms in the Criminal Justice System},
   institution = {Electronic Privacy Information Centre},
   url = {https://epic.org/algorithmic-transparency/crim-justice/},
   type = {Report}
}

@article{Esteva17,
   author = {Esteva, Andre and Kuprel, Brett and Novoa, Roberto A. and Ko, Justin and Swetter, Susan M. and Blau, Helen M. and Thrun, Sebastian},
   year = {2017},
   title = {Dermatologist-level classification of skin cancer with deep neural networks},
   journaltitle = {Nature},
   volume = {542},
   pages = {115},
   DOI = {10.1038/nature21056},
   type = {Journal Article}
}

@book{Eubanks17,
   author = {Eubanks, Virginia},
   year = {2017},
   title = {Automating inequality : how high-tech tools profile, police, and punish the poor},
   publisher = {St. Martin's Press},
   location = {New York, NY},
   edition = {First Edition.},
   pages = {260 pages},
   ISBN = {9781250074317 (hardcover)},
   type = {Book}
}

@article{Feathers19,
   author = {Feathers, Todd},
   year = {2019},
   date = {20 Aug},
   title = {Flawed Algorithms Are Grading Millions of Students’ Essays},
   journaltitle = {Motherboard Tech by Vice},
   url = {https://www.vice.com/en\_us/article/pa7dj9/flawed-algorithms-are-grading-millions-of-students-essays},
   urldate = {2019-08-30},
   type = {Electronic Article}
}

@online{FederationJASON19,
   author = {{Federation of American Scientists}},
   year = {2019},
   date = {12 Mar},
   title = {{JASON} Defense Advisory Panel Reports},
   url = {https://fas.org/irp/agency/dod/jason/},
   urldate = {2019-04-05},
   type = {Web Page}
}

@article{Fraser18,
   author = {Fraser, Hamish and Coiera, Enrico and Wong, David},
   year = {2018},
   title = {Safety of patient-facing digital symptom checkers},
   journaltitle = {The Lancet},
   volume = {392},
   number = {10161},
   pages = {2263-2264},
   DOI = {10.1016/S0140-6736(18)32819-8},
   type = {Journal Article}
}

@report{Frontier18,
   author = {{Frontier Economics}},
   year = {2018},
   title = {The Impact of Artificial Intelligence on Work : An evidence review prepared for the Royal Society and the British Academy},
   url = {https://royalsociety.org/topics-policy/projects/ai-and-work/},
   type = {Report}
}

@inproceedings{Furman16,
   author = {Furman, Jason},
   year = {2016},
   title = {Is This Time Different? The Opportunities and Challenges of Artificial Intelligence},
   booktitle = {{AI} Now: The Social and Economic Implications of Artificial Intelligence Technologies in the Near Term},
   loction = {New York University, New York, NY},
   url = {https://obamawhitehouse.archives.gov/sites/default/files/page/files/20160707\_cea\_ai\_furman.pdf},
   type = {Conference Proceedings}
}

@article{Future15,
   author = {{Future of Life Institute}},
   year = {2015},
   title = {Autonomous Weapons: An Open Letter From {AI} \& Robotics Researchers},
   url = {https://futureoflife.org/open-letter-autonomous-weapons/},
   urldate = {2018-10-18},
   type = {Electronic Article}
}

@article{Garvie16,
   author = {Garvie, Clare and Frankle, Jonathan},
   year = {2016},
   date = {7 Apr},
   title = {Facial-Recognition Software Might Have a Racial Bias Problem},
   journaltitle = {The Atlantic},
   url = {https://www.theatlantic.com/technology/archive/2016/04/the-underlying-bias-of-facial-recognition-systems/476991/},
   urldate = {2019-04-04},
   type = {Electronic Article}
}

@unpublished{Gonzalez-Jimenez18,
   author = {González-Jiménez, Mario and Babayan, Simon A. and Khazaeli, Pegah and Doyle, Margaret and Walton, Finlay and Reedy, Elliott and Glew, Thomas and Viana, Mafalda and Ranford-Cartwright, Lisa and Niang, Abdoulaye and Siria, Doreen J. and Okumu, Fredros O. and Diabaté, Abdoulaye and Ferguson, Heather M. and Baldini, Francesco and Wynne, Klaas},
   year = {2018},
   title = {Prediction of malaria mosquito species and population age structure using mid-infrared spectroscopy and supervised machine learning},
   howpublished = {bioRxiv},
   DOI = {10.1101/414342},
   url = {http://biorxiv.org/content/early/2018/09/12/414342.abstract},
   type = {Preprint}
}

@article{Greenhalgh18,
   author = {Greenhalgh, Trisha and Thorne, Sally and Malterud, Kirsti},
   year = {2018},
   title = {Time to challenge the spurious hierarchy of systematic over narrative reviews?},
   journaltitle = {European Journal of Clinical Investigation},
   volume = {48},
   number = {6},
   pages = {e12931},
   DOI = {10.1111/eci.12931},
   type = {Journal Article}
}

@article{Grundy19,
   author = {Grundy, Quinn and Chiu, Kellia and Held, Fabian and Continella, Andrea and Bero, Lisa and Holz, Ralph},
   year = {2019},
   title = {Data sharing practices of medicines related apps and the mobile ecosystem: traffic, content, and network analysis},
   journaltitle = {BMJ},
   volume = {364},
   pages = {l920},
   DOI = {10.1136/bmj.l920},
   type = {Journal Article}
}

@article{Gulshan16,
   author = {Gulshan, V. and Peng, L. and Coram, M. and Stumpe, M. C. and Wu, D. and Narayanaswamy, A. and Venugopalan, S. and Widner, K. and Madams, T. and Cuadros, J. and Kim, R. and Raman, R. and Nelson, P. C. and Mega, J. L. and Webster, D. R.},
   year = {2016},
   title = {Development and Validation of a Deep Learning Algorithm for Detection of Diabetic Retinopathy in Retinal Fundus Photographs},
   journaltitle = {JAMA},
   volume = {316},
   number = {22},
   pages = {2402-2410},
   DOI = {10.1001/jama.2016.17216},
   type = {Journal Article}
}

@inbook{Hadjiisky17,
   author = {Hadjiisky, M. and Pal, L. A. and Walker, C.},
   year = {2017},
   title = {Introduction: traversing the terrain of policy transfer: theory, methods and overview},
   booktitle = {Public Policy Transfer},
   editor = {Hadjiisky, Magdaléna and Pal, Leslie A. and Walker, Christopher},
   publisher = {Edward Elgar Publishing},
   location = {Cheltenham, UK},
   chapter = {1},
   pages = {1–26},
   DOI = {https://doi.org/10.4337/9781785368042.00007},
   type = {Book Section}
}

@inproceedings{Harrington18,
   author = {Harrington, L. and Suriawinata, A. and MacKenzie, T. and Hassanpour, S.},
   year = {2018},
   title = {Application of machine learning on colonoscopy screening records for predicting colorectal polyp recurrence},
   booktitle = {IEEE International Conference on Bioinformatics and Biomedicine (BIBM)},
   pages = {993-998},
   DOI = {10.1109/BIBM.2018.8621455},
   type = {Conference Proceedings}
}

@article{Hern18,
   author = {Hern, Alex},
   year = {2018},
   date = {14 Nov},
   title = {Google 'betrays patient trust' with DeepMind Health move},
   journaltitle = {The Guardian},
   url = {https://www.theguardian.com/technology/2018/nov/14/google-betrays-patient-trust-deepmind-healthcare-move},
   urldate = {2019-08-11},
   type = {Electronic Article}
}

@inproceedings{Hibbert18,
   author = {Hibbert, Marienne and Georgeff, Michael},
   year = {2018},
   title = {Health Care Home Risk Stratification Tool},
   booktitle = {HIC 2018},
   loction = {Sydney, Australia},
   url = {https://www.hisa.org.au/slides/hic18/mon/MarienneHibbert.pdf},
   type = {Conference Proceedings}
}

@report{Holland18,
   author = {Holland, Chris},
   year = {2018},
   title = {2018 Close The Gap. A ten-year review: the Closing the Gap Strategy and Recommendations for Reset},
   institution = {Close the Gap Campaign Steering Committee for Indigenous Health Equality},
   url = {https://www.humanrights.gov.au/our-work/aboriginal-and-torres-strait-islander-social-justice/projects/close-gap-indigenous-health#EwVHY},
   type = {Report}
}

@report{HouseLords18Ready,
   author = {{House of Lords Select Committee on Artificial Intelligence}},
   year = {2018},
   title = {{AI} in the {UK}: ready, willing and able?},
   institution = {United Kingdom Parliament},
   url = {https://www.parliament.uk/ai-committee},
   type = {Report}
}

@inproceedings{Huang18,
   author = {Huang, C. Y. and Yang, M. C. and Huang, C. Y. and Chen, Y. J. and Wu, M. L. and Chen, K. W.},
   year = {2018},
   title = {A Chatbot-supported Smart Wireless Interactive Healthcare System for Weight Control and Health Promotion},
   booktitle = {IEEE International Conference on Industrial Engineering and Engineering Management},
   pages = {1791-1795},
   url = {<Go to ISI>://WOS:000458674600357},
   type = {Conference Proceedings}
}

@report{IPSOS16,
   author = {{Ipsos MORI Social Research Institute}},
   year = {2016},
   title = {The One-Way Mirror: Public attitudes to commercial access to health data},
   institution = {Report prepared for the Wellcome Trust},
   url = {https://wellcome.ac.uk/sites/default/files/public-attitudes-to-commercial-access-to-health-data-wellcome-mar16.pdf},
   type = {Report}
}

@article{Jain19,
   author = {Jain, Tara and Schwarz, Eleanor B. and Mehrotra, Ateev},
   year = {2019},
   title = {A Study of Telecontraception},
   journaltitle = {New England Journal of Medicine},
   volume = {381},
   number = {13},
   pages = {1287-1288},
   DOI = {10.1056/NEJMc1907545},
   type = {Journal Article}
}

@article{Jasanoff17,
   author = {Jasanoff, Sheila},
   year = {2017},
   title = {Virtual, visible, and actionable: Data assemblages and the sightlines of justice},
   journaltitle = {Big Data \& Society},
   volume = {4},
   number = {2},
   pages = {1-15},
   DOI = {10.1177/2053951717724477},
   type = {Journal Article}
}

@article{KamelBoulos19,
   author = {Kamel Boulos, M. N. and Peng, G. and VoPham, T.},
   year = {2019},
   title = {An overview of GeoAI applications in health and healthcare},
   journaltitle = {Int J Health Geogr},
   volume = {18},
   number = {1},
   pages = {7},
   DOI = {10.1186/s12942-019-0171-2},
   type = {Journal Article}
}

@inproceedings{Katsis17,
   author = {Katsis, Y. and Balac, N. and Chapman, D. and Kapoor, M. and Block, J. and Griswold, W. G. and Huang, J. and Koulouris, N. and Menarini, M. and Nandigam, V. and Ngo, M. and Ong, K. W. and Papakonstantinou, Y. and Smith, B. and Zarifis, K. and Woolf, S. and Patrick, K.},
   year = {2017},
   title = {Big Data Techniques for Public Health: A Case Study},
   booktitle = {IEEE/ACM International Conference on Connected Health: Applications, Systems and Engineering Technologies (CHASE)},
   pages = {222-231},
   DOI = {10.1109/CHASE.2017.81},
   type = {Conference Proceedings}
}

@inproceedings{Khourdifi18,
   author = {Khourdifi, Y. and Bahaj, M.},
   year = {2018},
   title = {Applying Best Machine Learning Algorithms for Breast Cancer Prediction and Classification},
   booktitle = {International Conference on Electronics, Control, Optimization and Computer Science (ICECOCS)},
   pages = {1-5},
   DOI = {10.1109/ICECOCS.2018.8610632},
   type = {Conference Proceedings}
}

@article{Khoury16,
   author = {Khoury, Muin J. and Galea, Sandro},
   year = {2016},
   title = {Will Precision Medicine Improve Population Health?},
   journaltitle = {JAMA},
   volume = {316},
   number = {13},
   pages = {1357-1358},
   DOI = {10.1001/jama.2016.12260},
   type = {Journal Article}
}

@article{Khoury14,
   author = {Khoury, Muin J. and Ioannidis, J. P. A.},
   year = {2014},
   title = {Big data meets public health},
   journaltitle = {Science},
   volume = {346},
   number = {6213},
   pages = {1054-1055},
   DOI = {10.1126/science.aaa2709},
   type = {Journal Article}
}

@article{Knight17,
   author = {Knight, Will},
   year = {2017},
   date = {11 Apr},
   title = {The Dark Secret at the Heart of {AI}},
   journaltitle = {MIT Technology Review},
   url = {https://www.technologyreview.com/s/604087/the-dark-secret-at-the-heart-of-ai/},
   urldate = {2018-10-25},
   type = {Electronic Article}
}

@article{Krieger12,
   author = {Krieger, N.},
   year = {2012},
   title = {Who and what is a "population"? Historical debates, current controversies, and implications for understanding "population health" and rectifying health inequities},
   journaltitle = {Milbank Q},
   volume = {90},
   number = {4},
   pages = {634-81},
   DOI = {10.1111/j.1468-0009.2012.00678.x},
   type = {Journal Article}
}

@article{Krieger17,
   author = {Krieger, N.},
   year = {2017},
   title = {Health Equity and the Fallacy of Treating Causes of Population Health as if They Sum to 100},
   journaltitle = {Am J Public Health},
   volume = {107},
   number = {4},
   pages = {541-549},
   DOI = {10.2105/ajph.2017.303655},
   type = {Journal Article}
}

@book{Kurzweil05,
   author = {Kurzweil, Ray},
   year = {2005},
   title = {The singularity is near : when humans transcend biology},
   publisher = {Viking},
   location = {New York},
   pages = {xvii, 652 p.},
   ISBN = {0670033847 (hardcover)},
   type = {Book}
}

@article{Kwon18,
   author = {Kwon, J. M. and Lee, Y. and Lee, Y. and Lee, S. and Park, J.},
   year = {2018},
   title = {An Algorithm Based on Deep Learning for Predicting In-Hospital Cardiac Arrest},
   journaltitle = {J Am Heart Assoc},
   volume = {7},
   number = {13},
   DOI = {10.1161/jaha.118.008678},
   type = {Journal Article}
}

@article{Labovitz17,
   author = {Labovitz Daniel, L. and Shafner, Laura and Reyes Gil, Morayma and Virmani, Deepti and Hanina, Adam},
   year = {2017},
   title = {Using Artificial Intelligence to Reduce the Risk of Nonadherence in Patients on Anticoagulation Therapy},
   journaltitle = {Stroke},
   volume = {48},
   number = {5},
   pages = {1416-1419},
   DOI = {10.1161/STROKEAHA.116.016281},
   type = {Journal Article}
}

@article{Lake19,
   author = {Lake, I. R. and Colon-Gonzalez, F. J. and Barker, G. C. and Morbey, R. A. and Smith, G. E. and Elliot, A. J.},
   year = {2019},
   title = {Machine learning to refine decision making within a syndromic surveillance service},
   journaltitle = {BMC Public Health},
   volume = {19},
   number = {1},
   pages = {559},
   DOI = {10.1186/s12889-019-6916-9},
   type = {Journal Article}
}

@unpublished{Lambrecht18,
   author = {Lambrecht, Anja and Tucker, Catherine E.},
   year = {2018},
   title = {Algorithmic Bias? An Empirical Study into Apparent Gender-Based Discrimination in the Display of STEM Career Ads},
   howpublished = {SSRN},
   url = {https://dx.doi.org/10.2139/ssrn.2852260},
   type = {Preprint}
}

@article{Lapowsky18,
   author = {Lapowsky, Issie},
   year = {2018},
   date = {17 Jan},
   title = {Crime-Predicting Algorithms May Not Fare Much Better Than Untrained Humans},
   journaltitle = {Wired Magazine},
   url = {https://www.wired.com/story/crime-predicting-algorithms-may-not-outperform-untrained-humans/},
   urldate = {2018-12-10},
   type = {Electronic Article}
}

@article{Lavanchy18,
   author = {Lavanchy, Maude},
   year = {2018},
   date = {1 Nov},
   title = {Amazon’s sexist hiring algorithm could still be better than a human},
   journaltitle = {The Conversation},
   url = {https://theconversation.com/amazons-sexist-hiring-algorithm-could-still-be-better-than-a-human-105270},
   urldate = {2018-11-16},
   type = {Electronic Article}
}

@article{Lavigne19,
   author = {Lavigne, M. and Mussa, F. and Creatore, M. I. and Hoffman, S. J. and Buckeridge, D. L.},
   year = {2019},
   title = {A population health perspective on artificial intelligence},
   journaltitle = {Healthc Manage Forum},
   volume = {32},
   number = {4},
   pages = {173-177},
   DOI = {10.1177/0840470419848428},
   type = {Journal Article}
}

@article{LePage19,
   author = {Le Page, Michael},
   year = {2019},
   date = {26 Jun},
   title = {AIs that diagnose diseases are starting to assist and replace doctors},
   journaltitle = {New Scientist},
   url = {https://www.newscientist.com/article/mg24232363-000-ais-that-diagnose-diseases-are-starting-to-assist-and-replace-doctors/},
   urldate = {2019-08-23},
   type = {Electronic Article}
}

@article{Lecher18,
   author = {Lecher, Colin},
   year = {2018},
   date = {21 Mar},
   title = {What Happens When an Algorithm Cuts Your Healthcare},
   journaltitle = {The Verge},
   url = {https://www.theverge.com/2018/3/21/17144260/healthcare-medicaid-algorithm-arkansas-cerebral-palsy},
   urldate = {2019-08-11},
   type = {Electronic Article}
}

@article{LeCun15,
   author = {LeCun, Yann and Bengio, Yoshua and Hinton, Geoffrey},
   year = {2015},
   title = {Deep learning},
   journaltitle = {Nature},
   volume = {521},
   pages = {436},
   DOI = {10.1038/nature14539},
   type = {Journal Article}
}

@article{Leonelli19,
   author = {Leonelli, Sabina},
   year = {2019},
   date = {15 Oct},
   title = {Data — from objects to assets},
   journaltitle = {Nature},
   url = {https://www.nature.com/articles/d41586-019-03062-w},
   urldate = {2019-10-25},
   type = {Electronic Article}
}

@inproceedings{Li17DeepAirPollution,
   author = {Li, V. O. K. and Lam, J. C. K. and Chen, Y. and Gu, J.},
   year = {2017},
   title = {Deep Learning Model to Estimate Air Pollution Using M-BP to Fill in Missing Proxy Urban Data},
   booktitle = {GLOBECOM 2017 - IEEE Global Communications Conference},
   pages = {1-6},
   DOI = {10.1109/GLOCOM.2017.8255004},
   type = {Conference Proceedings}
}

@article{Lim17,
   author = {Lim, S. and Tucker, C. S. and Kumara, S.},
   year = {2017},
   title = {An unsupervised machine learning model for discovering latent infectious diseases using social media data},
   journaltitle = {Journal of Biomedical Informatics},
   volume = {66},
   pages = {82-94},
   DOI = {10.1016/j.jbi.2016.12.007},
   type = {Journal Article}
}

@article{Lohr18,
   author = {Lohr, Steve},
   year = {2018},
   date = {9 Feb},
   title = {Facial Recognition Is Accurate, if You’re a White Guy},
   journaltitle = {The New York Times},
   url = {https://www.nytimes.com/2018/02/09/technology/facial-recognition-race-artificial-intelligence.html},
   urldate = {2019-04-04},
   type = {Electronic Article}
}

@article{Lomas18,
   author = {Lomas, Natasha},
   year = {2018},
   date = {13 Jun},
   title = {Audit of NHS Trust’s app project with DeepMind raises more questions than it answers},
   journaltitle = {TechCrunch},
   url = {https://techcrunch.com/2018/06/13/audit-of-nhs-trusts-app-project-with-deepmind-raises-more-questions-than-it-answers/},
   urldate = {2019-01-25},
   type = {Electronic Article}
}

@article{Luckin17,
   author = {Luckin, Rose},
   year = {2017},
   title = {Towards artificial intelligence-based assessment systems},
   journaltitle = {Nature Human Behaviour},
   volume = {1},
   pages = {0028},
   DOI = {10.1038/s41562-016-0028},
   type = {Journal Article}
}

@article{Lum16,
   author = {Lum, Kristian and Isaac, William},
   year = {2016},
   title = {To predict and serve?},
   journaltitle = {Significance},
   volume = {13},
   number = {5},
   pages = {14-19},
   DOI = {10.1111/j.1740-9713.2016.00960.x},
   type = {Journal Article}
}

@article{Lunden19,
   author = {Lunden, Ingrid},
   year = {2019},
   date = {2 Aug},
   title = {Babylon Health confirms \$550M raise at \$2B+ valuation to expand its {AI}-based health services},
   journaltitle = {TechCrunch},
   url = {https://techcrunch.com/2019/08/02/babylon-health-confirms-550m-raise-to-expand-its-ai-based-health-services-to-the-us-and-asia/},
   urldate = {2019-08-30},
   type = {Electronic Article}
}

@article{Mackenzie15,
   author = {Mackenzie, Adrian},
   year = {2015},
   title = {The production of prediction: What does machine learning want?},
   journaltitle = {European Journal of Cultural Studies},
   volume = {18},
   number = {4-5},
   pages = {429-445},
   DOI = {10.1177/1367549415577384},
   type = {Journal Article}
}

@article{Maddox19,
   author = {Maddox, Thomas M. and Rumsfeld, John S. and Payne, Philip R. O.},
   year = {2019},
   title = {Questions for Artificial Intelligence in Health Care},
   journaltitle = {JAMA},
   volume = {321},
   number = {1},
   pages = {31-32},
   DOI = {10.1001/jama.2018.18932},
   type = {Journal Article}
}

@inproceedings{Madnani17,
   author = {Madnani, Nitin and Loukina, Anastassia and von Davier, Alina and Burstein, Jill and Cahill, Aoife},
   year = {2017},
   title = {Building Better Open-Source Tools to Support Fairness in Automated Scoring},
   booktitle = {First ACL Workshop on Ethics in Natural Language Processing},
   loction = {Valencia, Spain},
   publisher = {Association for Computational Linguistics},
   pages = {41-52},
   DOI = {https://doi.org/10.18653/v1/W17-1605},
   url = {https://doi.org/10.18653/v1/W17-1605},
   type = {Conference Proceedings}
}

@article{Mann16,
   author = {Mann, Gideon and Cathy, O'Neil.},
   year = {2016},
   date = {9 Dec},
   title = {Hiring Algorithms Are Not Neutral},
   journaltitle = {Harvard Business Review},
   url = {https://hbr.org/2016/12/hiring-algorithms-are-not-neutral},
   urldate = {2019-05-16},
   type = {Electronic Article}
}

@article{Marmot05,
   author = {Marmot, M.},
   year = {2005},
   title = {Social determinants of health inequalities},
   journaltitle = {Lancet},
   volume = {365},
   number = {9464},
   pages = {1099-104},
   DOI = {10.1016/S0140-6736(05)71146-6},
   type = {Journal Article}
}

@article{Mattick14,
   author = {Mattick, J. S. and Dziadek, M. A. and Terrill, B. N. and Kaplan, W. and Spigelman, A. D. and Bowling, F. G. and Dinger, M. E.},
   year = {2014},
   title = {The impact of genomics on the future of medicine and health},
   journaltitle = {Med J Aust},
   volume = {201},
   number = {1},
   pages = {17-20},
   type = {Journal Article}
}

@article{McGoey17,
   author = {McGoey, Linsey},
   year = {2017},
   title = {The Elusive Rentier Rich: Piketty’s Data Battles and the Power of Absent Evidence},
   journaltitle = {Science, Technology, \& Human Values},
   volume = {42},
   number = {2},
   pages = {257-279},
   DOI = {10.1177/0162243916682598},
   type = {Journal Article}
}

@report{McMorrow14,
   author = {McMorrow, Dan},
   year = {2014},
   title = {Data for Individual Health. Report prepared for Agency for Healthcare Research and Quality},
   number = {JSR-14-TASK-007},
   institution = {{JASON} The MITRE Corporation},
   url = {https://fas.org/irp/agency/dod/jason/data-health.pdf},
   type = {Report}
}

@article{Meagher17,
   author = {Meagher, K. M. and McGowan, M. L. and Settersten, R. A., Jr. and Fishman, J. R. and Juengst, E. T.},
   year = {2017},
   title = {Precisely Where Are We Going? Charting the New Terrain of Precision Prevention},
   journaltitle = {Annu Rev Genomics Hum Genet},
   volume = {18},
   pages = {369-387},
   DOI = {10.1146/annurev-genom-091416-035222},
   type = {Journal Article}
}

@article{Medhora19,
   author = {Medhora, Shalailah},
   year = {2019},
   date = {18 Feb},
   title = {Over 2000 people died after receiving Centrelink robo-debt notice, figures reveal},
   journaltitle = {ABC triple J Hack},
   url = {https://www.abc.net.au/triplej/programs/hack/2030-people-have-died-after-receiving-centrelink-robodebt-notice/10821272},
   urldate = {2019-04-14},
   type = {Electronic Article}
}

@article{MedTech18,
   author = {{MedTech Boston}},
   year = {2018},
   date = {5 Mar},
   title = {The Top 5 {AI} In Healthcare Startups},
   journaltitle = {MedTech Boston},
   url = {https://medtechboston.medstro.com/blog/2018/03/05/the-top-5-ai-in-healthcare-startups/},
   urldate = {2018-12-06},
   type = {Electronic Article}
}

@article{Mesko19,
   author = {Meskó, Bertalan},
   year = {2019},
   date = {6 Jun},
   title = {FDA Approvals For Smart Algorithms In Medicine In One Giant Infographic},
   journaltitle = {The Medical Futurist},
   url = {https://medicalfuturist.com/fda-approvals-for-algorithms-in-medicine},
   urldate = {2019-08-29},
   type = {Electronic Article}
}

@article{Meyer11,
   author = {Meyer, Betrand},
   year = {2011},
   date = {28 Oct},
   title = {John McCarthy},
   journaltitle = {Communications of the ACM},
   url = {https://cacm.acm.org/blogs/blog-cacm/138907-john-mccarthy/fulltext},
   urldate = {2019-03-14},
   type = {Electronic Article}
}

@article{Miotto16,
   author = {Miotto, Riccardo and Li, Li and Kidd, Brian A. and Dudley, Joel T.},
   year = {2016},
   title = {Deep Patient: An Unsupervised Representation to Predict the Future of Patients from the Electronic Health Records},
   journaltitle = {Scientific Reports},
   volume = {6},
   pages = {26094},
   DOI = {10.1038/srep26094
https://www.nature.com/articles/srep26094#supplementary-information},
   type = {Journal Article}
}

@article{Mittelstadt2018,
   author = {Mittelstadt, Brent and Benzler, Justus and Engelmann, Lukas and Prainsack, Barbara and Vayena, Effy},
   year = {2018},
   title = {Is there a duty to participate in digital epidemiology?},
   journaltitle = {Life Sciences, Society and Policy},
   volume = {14},
   number = {1},
   pages = {9},
   DOI = {10.1186/s40504-018-0074-1},
   type = {Journal Article}
}

@article{Morcom18,
   author = {Morcom, Jake},
   year = {2018},
   date = {10 Sep},
   title = {Automating Inequality},
   journaltitle = {Think: Sustainability},
   editor = {Morcom, Jake},
   url = {https://2ser.com/automating-inequality/},
   note = {Podcast},
   type = {Podcast}
}

@article{Mukherjee17,
   author = {Mukherjee, Siddhartha},
   year = {2017},
   title = {{A.I. Versus M.D.}},
   journaltitle = {The New Yorker},
   volume = {Annals of Medicine},
   number = {April 3, 2017},
   url = {https://www.newyorker.com/magazine/2017/04/03/ai-versus-md},
   urldate = {2019-03-15},
   type = {Electronic Article}
}

@article{Muller19,
   author = {Muller, M. M. and Salathé, Marcel},
   year = {2019},
   title = {Crowdbreaks: Tracking Health Trends Using Public Social Media Data and Crowdsourcing},
   journaltitle = {Front Public Health},
   volume = {7},
   pages = {81},
   DOI = {10.3389/fpubh.2019.00081},
   type = {Journal Article}
}

@unpublished{Nadkarni19,
   author = {Nadkarni, Girish N. and Fleming, Fergus and McCullough, James R. and Chauhan, Kinsuk and Verghese, Divya A. and He, John C. and Quackenbush, John and Bonventre, Joseph V. and Murphy, Barbara and Parikh, Chirag R. and Donovan, Michael and Coca, Steven G.},
   year = {2019},
   title = {Prediction of rapid kidney function decline using machine learning combining blood biomarkers and electronic health record data},
   howpublished = {bioRxiv},
   DOI = {10.1101/587774},
   url = {http://biorxiv.org/content/early/2019/03/28/587774.abstract},
   type = {Preprint}
}

@inbook{NRCandIOM2013,
   author = {{National Research Council} and {Institute of Medicine}},
   year = {2013},
   title = {The National Academies Collection: Reports funded by National Institutes of Health},
   booktitle = {U.S. Health in International Perspective: Shorter Lives, Poorer Health},
   editor = {Woolf, S. H. and Aron, L.},
   publisher = {National Academies Press (US)
National Academy of Sciences.},
   location = {Washington (DC)},
   DOI = {10.17226/13497},
   type = {Book Section}
}

@article{Newell15,
   author = {Newell, Sue and Marabelli, Marco},
   year = {2015},
   title = {Strategic opportunities (and challenges) of algorithmic decision-making: A call for action on the long-term societal effects of ‘datification’},
   journaltitle = {The Journal of Strategic Information Systems},
   volume = {24},
   number = {1},
   pages = {3-14},
   DOI = {https://doi.org/10.1016/j.jsis.2015.02.001},
   type = {Journal Article}
}

@article{Nott17,
   author = {Nott, George},
   year = {2017},
   date = {5 Apr},
   title = {Genevieve Bell calls out the creeps and warns of analytics' unintended consequences},
   journaltitle = {CIO},
   url = {https://www.cio.com.au/article/617193/creepy-algorithms-make-humans-anxious-warns-leading-anthropologist/},
   urldate = {2018-08-12},
   type = {Electronic Article}
}

@inproceedings{Obermeyer19FAT,
   author = {Obermeyer, Ziad and Mullainathan, Sendhil},
   year = {2019},
   title = {Dissecting Racial Bias in an Algorithm that Guides Health Decisions for 70 Million People},
   booktitle = {Conference on Fairness, Accountability, and Transparency},
   loction = {Atlanta, GA, USA},
   publisher = {ACM},
   pages = {89-89},
   DOI = {10.1145/3287560.3287593},
   url = {https://dl.acm.org/citation.cfm?id=3287593},
   type = {Conference Paper}
}

@article{Obermeyer19,
   author = {Obermeyer, Ziad and Powers, Brian and Vogeli, Christine and Mullainathan, Sendhil},
   year = {2019},
   title = {Dissecting racial bias in an algorithm used to manage the health of populations},
   journaltitle = {Science},
   volume = {366},
   number = {6464},
   pages = {447},
   DOI = {10.1126/science.aax2342},
   type = {Journal Article}
}

@article{Ogino19,
   author = {Ogino, S. and Nowak, J. A. and Hamada, T. and Milner, D. A., Jr. and Nishihara, R.},
   year = {2019},
   title = {Insights into Pathogenic Interactions Among Environment, Host, and Tumor at the Crossroads of Molecular Pathology and Epidemiology},
   journaltitle = {Annu Rev Pathol},
   volume = {14},
   pages = {83-103},
   DOI = {10.1146/annurev-pathmechdis-012418-012818},
   type = {Journal Article}
}

@article{Ohm10,
   author = {Ohm, Paul},
   year = {2010},
   title = {Broken Promises of Privacy: Responding to the Surprising Failure of Anonymization},
   journaltitle = {UCLA Law Review},
   volume = {57},
   pages = {1701-1777},
   type = {Journal Article}
}

@article{Ornstein18,
   author = {Ornstein, Charles and Thomas, Katie},
   year = {2018},
   date = {20-Sep},
   title = {Sloan Kettering’s Cozy Deal With Start-Up Ignites a New Uproar},
   journaltitle = {The New York Times},
   url = {https://www.nytimes.com/2018/09/20/health/memorial-sloan-kettering-cancer-paige-ai.html},
   urldate = {2019-11-07},
   type = {Electronic Article}
}

@report{OVIC18,
   author = {OVIC},
   year = {2018},
   title = {Artificial intelligence and privacy Issues paper},
   institution = {Office of the Victorian Information Commissioner},
   url = {https://ovic.vic.gov.au/resource/artificial-intelligence-and-privacy/},
   type = {Report}
}

@report{Pamlin15,
   author = {Pamlin, Dennis and Armstrong, Stuart},
   year = {2015},
   title = {Global Challenges : 12 Risks that threaten human civilisation},
   institution = {Global Challenges Foundation},
   url = {https://api.globalchallenges.org/static/wp-content/uploads/12-Risks-with-infinite-impact.pdf},
   type = {Report}
}

@article{Panch19InconvientTruth,
   author = {Panch, Trishan and Mattie, Heather and Celi, Leo Anthony},
   year = {2019},
   title = {The “inconvenient truth” about {AI} in healthcare},
   journaltitle = {npj Digital Medicine},
   volume = {2},
   number = {1},
   pages = {77},
   DOI = {10.1038/s41746-019-0155-4},
   type = {Journal Article}
}

@article{Panch19AIOppRisks,
   author = {Panch, Trishan and Pearson-Stuttard, Jonathan and Greaves, Felix and Atun, Rifat},
   year = {2019},
   title = {Artificial intelligence: opportunities and risks for public health},
   journaltitle = {The Lancet Digital Health},
   volume = {1},
   number = {1},
   pages = {e13-e14},
   DOI = {https://doi.org/10.1016/S2589-7500(19)30002-0},
   type = {Journal Article}
}

@article{Park18,
   author = {Park, H. A. and Jung, H. and On, J. and Park, S. K. and Kang, H.},
   year = {2018},
   title = {Digital epidemiology: Use of digital data collected for non-epidemiological purposes in epidemiological studies},
   journaltitle = {Healthcare Informatics Research},
   volume = {24},
   number = {4},
   pages = {253-262},
   DOI = {10.4258/hir.2018.24.4.253},
   type = {Journal Article}
}

@inbook{Parry17,
   author = {Parry, Bronwyn and Greenhough, Beth},
   year = {2017},
   title = {Genesis: What is Bioinformation?},
   booktitle = {Bioinformation},
   publisher = {Cambridge, UK ; Malden, MA : Polity Press},
   chapter = {1},
   type = {Book Section}
}

@book{Pasquale15,
   author = {Pasquale, Frank},
   year = {2015},
   title = {The black box society : the secret algorithms that control money and information},
   publisher = {Harvard University Press},
   location = {Cambridge},
   pages = {311 pages},
   ISBN = {9780674368279},
   type = {Book}
}

@article{Pergialiotis18,
   author = {Pergialiotis, V. and Pouliakis, A. and Parthenis, C. and Damaskou, V. and Chrelias, C. and Papantoniou, N. and Panayiotides, I.},
   year = {2018},
   title = {The utility of artificial neural networks and classification and regression trees for the prediction of endometrial cancer in postmenopausal women},
   journaltitle = {Public Health},
   volume = {164},
   pages = {1-6},
   DOI = {10.1016/j.puhe.2018.07.012},
   type = {Journal Article}
}

@article{Perlman17,
   author = {Perlman, S. E. and McVeigh, K. H. and Thorpe, L. E. and Jacobson, L. and Greene, C. M. and Gwynn, R. C.},
   year = {2017},
   title = {Innovations in Population Health Surveillance: Using Electronic Health Records for Chronic Disease Surveillance},
   journaltitle = {Am J Public Health},
   volume = {107},
   number = {6},
   pages = {853-857},
   DOI = {10.2105/ajph.2017.303813},
   type = {Journal Article}
}

@inproceedings{Potash15,
   author = {Potash, Eric and Brew, Joe and Loewi, Alexander and Majumdar, Subhabrata and Reece, Andrew and Walsh, Joe and Rozier, Eric and Jorgenson, Emile and Mansour, Raed and Ghani, Rayid},
   year = {2015},
   title = {Predictive Modeling for Public Health: Preventing Childhood Lead Poisoning},
   booktitle = {21th ACM SIGKDD International Conference on Knowledge Discovery and Data Mining},
   loction = {Sydney, NSW, Australia},
   publisher = {ACM},
   pages = {2039-2047},
   DOI = {10.1145/2783258.2788629},
   type = {Conference Proceedings}
}

@article{Prainsack17,
   author = {Prainsack, Barbara},
   year = {2017},
   title = {The “We” in the “Me”: Solidarity and Health Care in the Era of Personalized Medicine},
   journaltitle = {Science, Technology, \& Human Values},
   volume = {43},
   number = {1},
   pages = {21-44},
   DOI = {10.1177/0162243917736139},
   type = {Journal Article}
}

@article{Prainsack19,
   author = {Prainsack, Barbara},
   year = {2019},
   title = {Precision Medicine Needs a Cure for Inequality},
   journaltitle = {Current History},
   volume = {118},
   number = {804},
   pages = {11-15},
   type = {Journal Article}
}

@unpublished{Prelot18,
   author = {Prélot, Laurie and Draisma, Harmen and Anasanti, Mila D. and Balkhiyarova, Zhanna and Wielscher, Matthias and Yengo, Loic and Balkau, Beverley and Roussel, Ronan and Sebert, Sylvain and Ala-Korpela, Mika and Froguel, Philippe and Jarvelin, Marjo-Riitta and Kaakinen, Marika and Prokopenko, Inga},
   year = {2018},
   title = {Machine Learning in Multi-Omics Data to Assess Longitudinal Predictors of Glycaemic Health},
   howpublished = {bioRxiv},
   DOI = {10.1101/358390},
   url = {http://biorxiv.org/content/early/2018/12/13/358390.abstract},
   type = {Preprint}
}

@article{PrivacyIntl16,
   author = {{Privacy International}},
   year = {2016},
   date = {1 Dec},
   title = {Algorithms, Intelligence, and Learning Oh My},
   url = {https://privacyinternational.org/feature/863/algorithms-intelligence-and-learning-oh-my},
   urldate = {2018-10-14},
   type = {Electronic Article}
}

@report{PrivacyIntl18,
   author = {{Privacy International} and {ARTICLE 19}},
   year = {2018},
   title = {Privacy and Freedom of Expression in the Age of Artificial Intelligence},
   institution = {Privacy International \& ARTICLE 19},
   url = {https://www.privacyinternational.org/topics/artificial-intelligence},
   type = {Report}
}

@article{Rahwan19,
   author = {Rahwan, Iyad and Cebrian, Manuel and Obradovich, Nick and Bongard, Josh and Bonnefon, Jean-François and Breazeal, Cynthia and Crandall, Jacob W. and Christakis, Nicholas A. and Couzin, Iain D. and Jackson, Matthew O. and Jennings, Nicholas R. and Kamar, Ece and Kloumann, Isabel M. and Larochelle, Hugo and Lazer, David and McElreath, Richard and Mislove, Alan and Parkes, David C. and Pentland, Alex ‘Sandy’ and Roberts, Margaret E. and Shariff, Azim and Tenenbaum, Joshua B. and Wellman, Michael},
   year = {2019},
   title = {Machine behaviour},
   journaltitle = {Nature},
   volume = {568},
   number = {7753},
   pages = {477-486},
   DOI = {10.1038/s41586-019-1138-y},
   type = {Journal Article}
}

@inproceedings{Rajiwall17,
   author = {Rajliwall, N. S. and Chetty, G. and Davey, R.},
   year = {2017},
   title = {Chronic disease risk monitoring based on an innovative predictive modelling framework},
   booktitle = {IEEE Symposium Series on Computational Intelligence (SSCI)},
   pages = {1-8},
   DOI = {10.1109/SSCI.2017.8285257},
   type = {Conference Proceedings}
}

@inproceedings{Rajiwall18,
   author = {Rajliwall, N. S. and Davey, R. and Chetty, G.},
   year = {2018},
   title = {Machine Learning Based Models for Cardiovascular Risk Prediction},
   booktitle = {International Conference on Machine Learning and Data Engineering (iCMLDE)},
   pages = {142-148},
   DOI = {10.1109/iCMLDE.2018.00034},
   type = {Conference Proceedings}
}

@article{Rajpurkar18,
   author = {Rajpurkar, Pranav and Irvin, Jeremy and Ball, Robyn L. and Zhu, Kaylie and Yang, Brandon and Mehta, Hershel and Duan, Tony and Ding, Daisy and Bagul, Aarti and Langlotz, Curtis P. and Patel, Bhavik N. and Yeom, Kristen W. and Shpanskaya, Katie and Blankenberg, Francis G. and Seekins, Jayne and Amrhein, Timothy J. and Mong, David A. and Halabi, Safwan S. and Zucker, Evan J. and Ng, Andrew Y. and Lungren, Matthew P.},
   year = {2018},
   title = {Deep learning for chest radiograph diagnosis: A retrospective comparison of the CheXNeXt algorithm to practicing radiologists},
   journaltitle = {PLOS Medicine},
   volume = {15},
   number = {11},
   pages = {e1002686},
   DOI = {10.1371/journal.pmed.1002686},
   type = {Journal Article}
}

@report{Reisman18AlgoImpactAssess,
   author = {Reisman, Dillon and Schultz, Jason and Crawford, Kate and Whittaker, Meredith},
   year = {2018},
   title = {Algorithmic Impact Assessments: A Practical Framework for Public Agency Accountability},
   institution = {{AI} Now Institute, New York University},
   url = {https://ainowinstitute.org/reports.html},
   type = {Report}
}

@article{Revell17,
   author = {Revell, Timothy},
   year = {2017},
   date = {16 May},
   title = {Google DeepMind NHS data deal was ‘legally inappropriate’},
   journaltitle = {New Scientist},
   url = {https://www.newscientist.com/article/2131256-google-deepmind-nhs-data-deal-was-legally-inappropriate/},
   urldate = {2019-03-22},
   type = {Electronic Article}
}

@unpublished{RichardsonForthcoming,
   author = {Richardson, Rashida and Schultz, Jason and Crawford, Kate},
   year = {forthcoming},
   title = {Dirty Data, Bad Predictions: How Civil Rights Violations Impact Police Data, Predictive Policing Systems, and Justice},
   howpublished = {New York University Law Review Online},
   url = {SSRN https://ssrn.com/abstract=3333423},
   type = {Preprint}
}

@report{Richmond19,
   author = {Richmond, Brigid},
   year = {2019},
   title = {A Day in the Life of Data: Removing the opacity surrounding the data collection, sharing and use environment in Australia},
   institution = {Consumer Policy Research Centre (CPRC)},
   url = {https://cprc.org.au/publication/research-report-a-day-in-the-life-of-data/},
   type = {Report}
}

@article{Rocher19,
   author = {Rocher, Luc and Hendrickx, Julien M. and de Montjoye, Yves-Alexandre},
   year = {2019},
   title = {Estimating the success of re-identifications in incomplete datasets using generative models},
   journaltitle = {Nature Communications},
   volume = {10},
   number = {1},
   pages = {3069},
   DOI = {10.1038/s41467-019-10933-3},
   type = {Journal Article}
}

@book{Rose08,
   author = {Rose, G. A. and Khaw, Kay-Tee and Marmot, M. G.},
   year = {2008},
   title = {Rose's strategy of preventive medicine : the complete original text},
   publisher = {Oxford University Press},
   location = {Oxford ; New York},
   edition = {New},
   pages = {xviii, 171 p.},
   ISBN = {9780192630971 (alk. paper)},
   url = {Table of contents http://www.loc.gov/catdir/toc/fy0804/2007045263.html},
   type = {Book}
}

@article{Rubens14,
   author = {Rubens, M. and Ramamoorthy, V. and Saxena, A. and Shehadeh, N.},
   year = {2014},
   title = {Public health in the twenty-first century: the role of advanced technologies},
   journaltitle = {Front Public Health},
   volume = {2},
   pages = {224},
   DOI = {10.3389/fpubh.2014.00224},
   type = {Journal Article}
}

@book{Russell10,
   author = {Russell, Stuart J. and Norvig, Peter},
   year = {2010},
   title = {Artificial intelligence : a modern approach},
   publisher = {Prentice Hall},
   location = {Upper Saddle River},
   edition = {3rd},
   series = {Prentice Hall series in artificial intelligence},
   pages = {xviii, 1132 p.},
   ISBN = {9780136042594
0136042597},
   type = {Book}
}
\end{document}